\documentclass[nohyperref]{article}

\usepackage{microtype}
\usepackage{graphicx}
\usepackage{subfigure}
\usepackage{booktabs} 
\usepackage[utf8]{inputenc} 
\usepackage[T1]{fontenc}    
\usepackage{booktabs}       
\usepackage{amsfonts}       
\usepackage{nicefrac}       
\usepackage{xcolor}         
\usepackage{siunitx}
\usepackage{amsmath}
\usepackage{amssymb}
\usepackage{mathtools}
\usepackage{amsthm}
\usepackage{float}
\usepackage{bbold}
\usepackage{placeins}
\usepackage{stfloats}
\usepackage[textsize=tiny]{todonotes}
\usepackage{siunitx}
\usepackage{hyperref}
\usepackage[capitalize,noabbrev]{cleveref}

\usepackage{icml2023_arxiv}

\usepackage{url}

\begin{document}

\twocolumn[
\icmltitle{Optical Transformers}

\icmlsetsymbol{equal}{*}

\begin{icmlauthorlist}
\icmlauthor{Maxwell G. Anderson}{aep}{,}
\icmlauthor{Shi-Yuan Ma}{aep}{,}
\icmlauthor{Tianyu Wang}{aep}{,}
\icmlauthor{Logan G. Wright}{aep,phi}{,}
\icmlauthor{Peter L. McMahon}{aep,kav}{}
\end{icmlauthorlist}

\icmlaffiliation{aep}{School of Applied and Engineering Physics, Cornell University, Ithaca, NY 14853, USA}
\icmlaffiliation{phi}{NTT Physics \& Informatics Laboratories, NTT Research, Inc., Sunnyvale, CA 94085, USA}
\icmlaffiliation{kav}{Kavli Institute at Cornell for Nanoscale Science, Cornell University, Ithaca, NY 14853, USA}

\icmlcorrespondingauthor{Peter L. McMahon}{pmcmahon@cornell.edu}
\icmlcorrespondingauthor{Maxwell G. Anderson}{mga58@cornell.edu}

\vskip 0.3in
]

\printAffiliationsAndNotice{}

\begin{abstract}
The rapidly increasing size of deep-learning models has caused renewed and growing interest in alternatives to digital electronic computers that could dramatically reduce the energy cost of running state-of-the-art neural networks. Optical matrix-vector multipliers are best suited to performing computations with very large operands, which suggests that large Transformer models could be a good target for optical computing. To test this idea, we performed small-scale optical experiments with a prototype accelerator to demonstrate that Transformer operations can run on optical hardware despite noise and errors. Using simulations, validated by our experiments, we then explored the energy efficiency of optical implementations of Transformers and identified scaling laws for model performance with respect to optical energy usage. We found that the optical energy per multiply-accumulate (MAC) scales as $\frac{1}{d}$ where $d$ is the Transformer width, an asymptotic advantage over the constant per-MAC energy use of digital systems. We conclude that with well-engineered, large-scale optical hardware, it may be possible to achieve a $100 \times$ energy-efficiency advantage for running some of the largest current Transformer models, and that if both the models and the optical hardware are scaled to the quadrillion-parameter regime, optical computers could have a $>$$8,000\times$ energy-efficiency advantage over state-of-the-art digital-electronic processors that achieve 300 fJ/MAC. We analyzed how these results motivate and inform the construction of future optical accelerators along with optics-amenable deep-learning approaches. With assumptions about future improvements to electronics and Transformer quantization techniques (5$\times$ cheaper memory access, double the digital--analog conversion efficiency, and 4-bit precision), we estimated that optical computers' advantage against current 300-fJ/MAC digital processors could grow to $>$$100,000\times$.
    
\end{abstract}

\section{Introduction}
Deep learning models' exponentially increasing scale is both a key driver in advancing the state-of-the-art and a cause of growing concern about the energy usage, speed, and therefore practicality of massive-scale deep learning. In the past few years, it has been found in particular that Transformer \cite{transformer} architectures significantly improve when sized up to billions or even trillions of parameters \cite{gpt3, kaplan_scaling_laws, routed_scaling_laws, chinchilla, efficient_scaling_survey, Zhai_2022_CVPR_scaling_vits}, causing an exponential growth of deep learning compute usage \cite{Sanh2019DistilBERTAD, large_scale_compute_trends}. These large-scale Transformers achieve ever more impressive results in not only natural language processing, but also in other domains such as computer vision \cite{ViT, liu2021Swin}, graphs \cite{kim2021transformers_graph}, and in multi-modal settings \cite{perceiver, perceiverIO, clip, dalle, coca, gato}, making them a popular but expensive solution for many tasks---digital hardware's energy efficiency (ie. per-flop or per-inference cost) has not kept up with the growing FLOP requirements of state-of-the-art deep learning models \cite{large_scale_compute_trends}. They also have transfer learning capabilities \cite{gpt, bert, gpt2, gpt3, lu2021fpt_universal_computation_xf, ViT}, allowing them to easily generalize to specific tasks, in some cases in a zero-shot setting where no further training is necessary \cite{gpt3, dalle, minerva}.

These models' cost and generic machine learning abilities strongly motivate the development of hardware accelerators for efficient and fast inference. In digital electronics, there has already been much effort in creating hardware for deep learning such as GPUs, mobile accelerator chips, FPGAs, and large-scale AI-dedicated accelerator systems \cite{reuthersurvey2020, graphcore, cerebras, nvidia_h100, habana}. But there is also growing interest in analog computing, where, among various approaches \cite{Sebastian2020}, optical neural networks (ONNs) have been proposed as solutions that may offer superior efficiency and latency over neural-network implementations on digital computers \cite{caulfield2010future, Wetzstein2020, nahmias2020photonic, stark2020opportunities, Huang2021Prospects, shastri2021photonics}. While these analog systems are subject to noise and error, neural network operations can be significantly cheaper when performed optically, often leaving the electrical overhead of loading the weights and data as the main cost, which is amortized in large linear operations. Ideally, the scaling is asymptotically better than digital systems in energy per MAC \cite{hamerly2019large, Wang2022, netcast, nahmias2020photonic}. Thus, acceleration of very wide, large-scale models---such as Transformers---is especially promising.

Here we show that Transformers increasingly leverage this scaling. To first demonstrate that Transformers may run on these systems despite their noise and error characteristics, we sampled operations from a real Transformer for language modelling to run on a real spatial light modulator (SLM) based experimental system, and used the results to create a calibrated simulation of a full Transformer running optically. We simulated with systematic error, noise, and imprecision of weights/inputs collected from these experiments and found that Transformers in these conditions still perform nearly as well as those running digitally. We summarize our key contributions as follows:
\begin{itemize}
    \item We demonstrated experimentally that linear operations in Transformers can be accurately run on real optical hardware, despite errors and noise.
    \item Via simulation, we established scaling laws for the performance and total energy costs of optical Transformers versus the model size and optical energy usage.
    \item We modeled the energy usage of a full ONN accelerator with a design based on our experiments and simulation.
    \item We estimated an orders-of-magnitude energy consumption advantage of optics versus state-of-the-art GPUs.
\end{itemize}
While our experiments and simulations were based on specific hardware as a representative example, our scope here is more general. We are interested in understanding how uniquely optical energy scaling and noise relate to Transformer performance and architecture. As such nearly all our findings apply broadly to linear optical processors, irrespective of their underlying hardware implementation details.

\section{Background and Related Work}
\subsection{Transformer Models}
Transformers are models for processing sequential data based on multi-head attention. Transformers consist of two-layer feed-forward blocks and multi-head attention (\cref{fig:xf}) operations. Multi-head attention computes relationships between sequence elements by deriving query, key, and value sequences $Q, K, V$ and computing dot products with a softmax nonlinearity in-between \cite{transformer}. Transformers also leverage modern design elements such as additive residual skip connections \cite{ResNet} and normalization layers \cite{layernorm}. A defining feature of Transformers is that entire sequences may be processed in matrix-matrix products in parallel (instead of one token/input at a time).

\subsection{Optical Accelerators}
There are myriad optical accelerator designs for deep learning. Despite their differences, the systems, including ours, have the following typical traits:

\begin{figure}[t]
  \centering
  \includegraphics[width=0.47\textwidth]{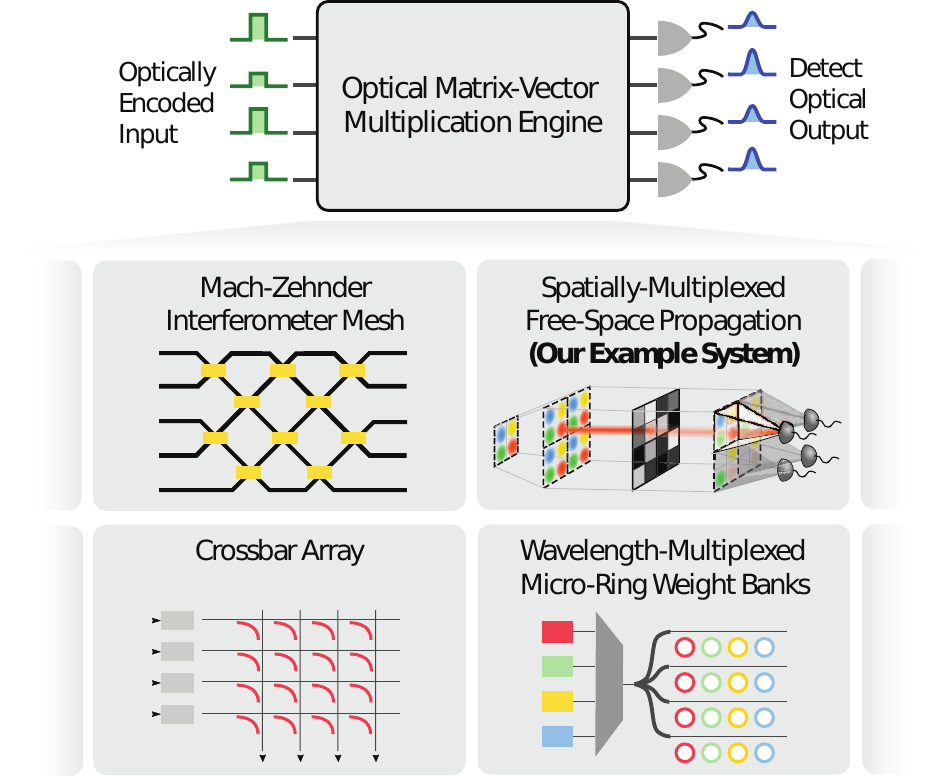}
  \caption{\textbf{General scheme of an optical neural network (ONN) accelerator.} Data is encoded and fed into the network, and the output is subject to shot noise. There are many experimental realizations of ONN accelerators such as Mach-Zehnder Interferometer meshes \cite{shen2017deep, bogaerts2020programmable}, crossbar arrays \cite{mrr_crossbars,feldmann2020parallel}, and wavelength-multiplexed micro-ring weight banks \cite{Tait:15}. In this work, we adopt the free-space multiplier \cite{Wang2022, spall2020fully, hayasaki1992optical} (top right) to demonstrate Transformer operations in optical experiments and as an example for our simulations, but our findings about Transformers on optical systems apply broadly to many optical-accelerator architectures, including those depicted in the inset.}
  \label{fig:intro}
\end{figure}

\paragraph{Acceleration of Linear Operations} Researchers have explored a wide variety of controllable optical systems (\cref{fig:intro}) which manipulate different types of optical modes to effectively implement arbitrary matrix-vector multiplications, vector-vector dot products \cite{shen2017deep, andregg2018wavelength, hamerly2019large, spall2020fully, bogaerts2020programmable, Wang2022}, or convolutions \cite{wu2020programmable, feldmann2020parallel, miscuglio2020massively, xu202111}. These linear operations can be performed efficiently and are the bulk of computations needed in neural networks.

\paragraph{Device Imprecision and Optical Shot Noise} Optical systems are subject to noise in both the actual hardware and from photon detection. Detection of optical intensity in particular is subject to a phenomenon known as \textit{shot noise} where the detected value is Poisson distributed. Notably, shot noise scales with the number of photons at the output of a dot product operation (ie. the accumulated sum of element-wise products): given vectors $x$ and $w$, with the elements of $x$ encoded as optical intensity, the output $Y$ is distributed as:

\begin{equation}
\label{eq:poisson}
    Y \sim \textrm{Poisson}(w \cdot x)
\end{equation}

Thus, optical neural networks are subject to noise which effectively limits the number of distinguishable output levels based on the number of photons used. For other vector or weight encoding schemes such as amplitude or phase encoding, equation \ref{eq:poisson} should be modified, but the detection is still subject to shot noise.

\paragraph{Efficient Photon Usage} 
Shot noise, and therefore an optical dot product's signal-to-noise ratio (SNR, which serves as an effective bit precision) is related to the mean number of photons at the \textit{output}. The efficiency of photon usage therefore grows with increasing multiply-accumulate operations (MACs): ideally, if the size of a dot product doubles, then each element needs half the photons for roughly constant output precision. The SNR for the Poisson distribution is given by

\begin{equation}
    \textrm{SNR}(Y) = \frac{\textrm{E}[Y]}{\sqrt{\textrm{Var}[Y]}} = \sqrt{w \cdot x} = \sqrt{\textrm{E}[Y]}
\end{equation}

which explains this behavior: if the expected size of the output $Y$ is held constant, then the SNR is the same, regardless of the dot product size.

Work on ONNs has characterized behavior in various scenarios by considering large operations and the resultant scaling \cite{hamerly2019large, nahmias2020photonic, Wang2022, netcast}. This efficient scaling is not a guarantee---the required number of photons may be influenced by a model architecture's activation/weight distributions, encoding schemes, precision requirements, etc.

\paragraph{Optical Neural Network Energy Costs} 
\label{par:energy}
The energy cost of optical neural networks is broken down into the optical costs of performing MACs and the electrical costs of loading/detecting data, which are usually dominant. Consider a product between two matrices, $A \in \mathbb{R}^{n \times d}$, $B \in \mathbb{R}^{d \times k}$. Such a product results in loading (detecting) $nd + dk$ ($nk$) scalars, and performing $ndk$ MACs. If the energy to electrically load (detect) a scalar is $E_{\mathrm{load}}$ ($E_{\mathrm{det}}$), and to perform a MAC optically is $E_{\mathrm{optical}}$, then the total energy is:

\begin{equation}
\label{eq:energy}
    E = (nd + dk)E_{\mathrm{load}} + nkE_{\mathrm{det}} + ndkE_{\mathrm{optical}}
\end{equation}

When both are present, efficient data transport and photon scaling cause optical systems to be asymptotically more efficient than digital electronic ones. Notice that regardless of the number of reuses, all data is only loaded once in Equation \ref{eq:energy}. That is, a vector only needs to be loaded once to be multiplied with many rows of a weight matrix. This is because copying a vector's data and transporting it is free optically. What remains is the amount of optical intensity to encode all of the copies, $E_{\mathrm{optical}}$, which ideally scales as $1/d$. These properties make energy cost disproportional to the number of MACs, $ndk$. This is an asymptotic advantage over electronic systems: $\frac{E_{\mathrm{digital}}}{E_{\mathrm{ONN}}} \sim \textrm{min}(n, d)$.

\paragraph{Streaming Weights Versus Weights-In-Place} \label{ss:wip} There are two approaches for loading weights.\textit{Weights-in-place} schemes involve loading them once, and re-using them for many inputs. Alternatively, systems can employ \textit{streaming weights} where at every computation the required weight matrix is loaded. Our experimental system is a weights-in-place scheme. For weights-in-place operations, the advantage scales as just $\frac{E_{\mathrm{digital}}}{E_{\mathrm{ONN}}} \sim d$.

\subsection{Previous Optical Neural Network Architectures}
Previous work has considered deep learning models such as MLPs and convolutional networks on benchmark tasks like MNIST \cite{miscuglio2020massively, Wang2022}, and simulations of larger convolutional models such as AlexNet \cite{alexnet} on more difficult datasets such as ImageNet \cite{hamerly2019large}. This begs the question of how well newer, larger models perform on optical systems.

\section{Optical Transformers}

\begin{figure*}
  \centering
  \includegraphics[width=0.95\textwidth]{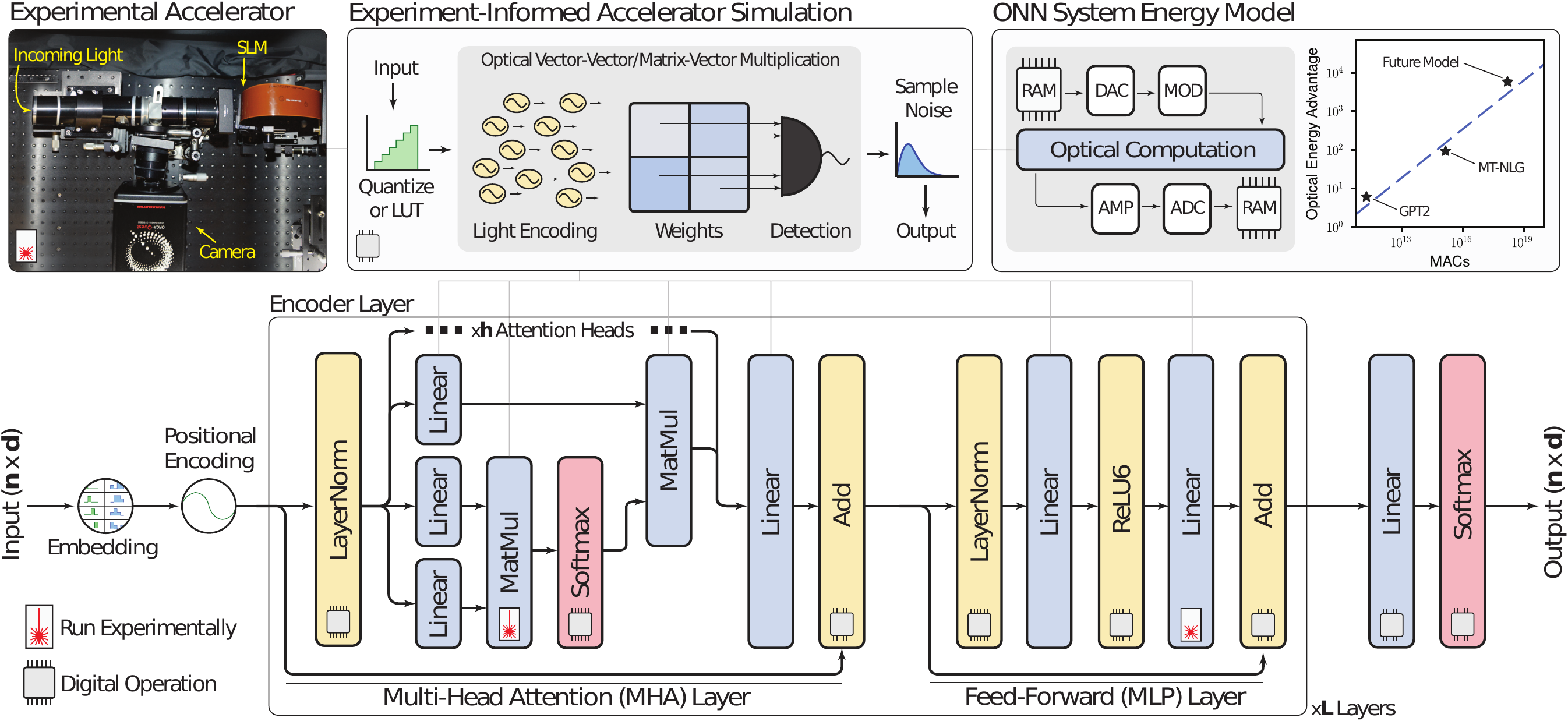}
  \caption{\textbf{Optical Transformer evaluation: prototype hardware; simulator model; Transformer architecture.} Bottom: typical Transformer architecture, but with ReLU6 activation. Top Left: experimental spatial light modulator (SLM)-based accelerator setup. From some layers---marked with a laser icon---we sampled dot products to run on real hardware. Top Middle: Linear operations, in light blue, run on a simulated accelerator with noise/error. Lookup tables (LUT) allow simulation using our setup's supported weight/activation values. Top right: our model of energy consumption for optical accelerators, based on assumptions and results from our experiment/simulations. The model accelerator system consists of random-access memory (RAM), a digital--analog converter (DAC), light modulation (MOD), amplification (AMP), an analog--digital converter (ADC), and an optical component that performs the computations.}
  \label{fig:xf}
\end{figure*}

We designed models that are intentionally similar to other Transformers, with the goal of simulating their behavior (informed by some experimental measurements) and energy consumption on optical hardware. A summary of our approach and model is in \cref{fig:xf}.

\subsection{Architecture and Task}
We created optical Transformer models with a GPT2-like \cite{gpt2} architecture that replaces the GELU \cite{hendrycks2016gelu} activation with ReLU6, which is known to improve low-precision model performance \cite{relu6, Howard2017MobileNetsEC, bnn_activations}. For language modelling, we used the raw Wikitext-103 dataset \cite{wikitext_103}. The models we simulated have 12 layers (consisting of multi-head attention and feed-forward blocks), operate on a context length of 1024 tokens, use 12 attention heads, and have embedding dimension $d$ varying from 192 to 1536. The full details of the training technique, architecture, and hyperparameters are in Appendix \ref{appendix:training}.

\subsection{Transformer Computations on Optical Hardware}
\label{ss:expt}
We adopt as a representative example of an optical accelerator a spatial light modulator (SLM) based system which computes vector-vector dot products \cite{Wang2022}. Vectors are encoded on a display, and copies are shone through the SLM which has varying transmission corresponding to some data (ie. a weight matrix). The outputs of this operation---element-wise products---are collected at detectors as the resultant dot products (\cref{fig:intro}, top right). 

We ran experiments using a real Transformer's (we used the base-sized model with $d=768$) weights in order to characterize the behavior of the system. We collected lookup tables (LUTs)---mappings of how input levels affect the light---and used them to train a ``LUT-aware'' optical Transformer model to run on the setup. Because our experimental setup is incapable of running the full model due to size/time constraints, we instead sampled vector-vector dot products from different parts of the model (marked with laser icon, \cref{fig:xf}). This allowed experiments to be run on data with the statistics of a typical model. The resultant noise model, lookup tables, and other details of the setup were used to build a calibrated, accurate simulation of the system, on which we ran full Transformer models and compared behavior with the experiment. The full details of the experiment, dot product sampling, and calibration are in Appendix \ref{appendix:expt}. 

\subsection{Simulation of Transformers on Optical Hardware}
\label{ss:sim}
Informed by our experiments, we constructed simulations of Transformers running on an optical accelerator. In particular we aimed to emulate the noise, error, and precision that we observed in order to understand how well full Transformers would perform when running on optical hardware.
\label{sec:sim}

\paragraph{Hybrid Scheme} Pure optical systems cannot easily compute activation or normalization functions. Thus we assumed they are computed digitally, and only simulated linear operations optically. Fortunately, the bulk of Transformer computations are within the layers that can be accelerated optically; even in smaller models, linear computations are the overwhelming majority (\cref{sec:energy_analysis}).

\paragraph{Non-Negative Weights and Inputs} An important limitation is that the display and SLM only support positive values. We worked around this by decomposing products into sums/differences of products with non-negative operands. Consider a product between matrices $W$ and $X$. If we let $W_+$ ($X_+$) and $W_-$ ($X_-$) be matrices with only the positive and negative elements of $W$ ($X$) respectively, then:

\begin{equation} \label{eq:quadruplePass}
   WX = W_+X_+ - |W_-|X_+ - W_+|X_-| + W_-X_-
\end{equation}

The constraint of having all-positive data is present in many but not all optical neural network systems.

\paragraph{Device Quantization} Actual hardware in an optical setup may only have certain number of representable levels. To emulate this behavior, relevant quantization was applied to the input and weights of the simulated model, and we trained the model using quantization-aware training \cite{jacob2018quantization}(QAT). For models to run on hardware we used the LUTs. We trained with 8-bit quantization and kept an exponential moving average (EMA) of activation min/max values, or percentile-clipped them. The full details of the scheme are in Appendix \ref{appendix:training}.

\paragraph{Systematic Errors}
Real systems may also have systematic errors that limit their effective precision. These are issues like cross-talk or misalignment in spatially multiplexed systems like ours, defects in chip elements in integrated photonic systems, etc. We simulated such a constraint by adding Gaussian noise to simulated model outputs (\cref{fig:xf}), scaled relative to the mean sizes of the outputs, as this was the noise behavior we observed in the experimental system. For example, we might simulate the Transformer with noise at each output equal to 5\% of the mean output value.

\paragraph{Optical Encoding and Shot Noise}
To simulate the optical computations we encoded the values of activation data as mean optical intensity. The outputs of our model were then subject to simulated shot noise (\cref{fig:xf}), which differs from the ``scale-relative-gaussian'' systematic error. To do this, outputs were scaled by a number such that the average photon number per feature (photons/MAC) was some target value. Each of these features was used as the mean of a Poisson distribution, from which a sample was drawn. The sampled outputs were then scaled back down to represent neuron values.

\section{Results}

\begin{figure}[t!]
    \centering
    \includegraphics[width=0.75\textwidth]{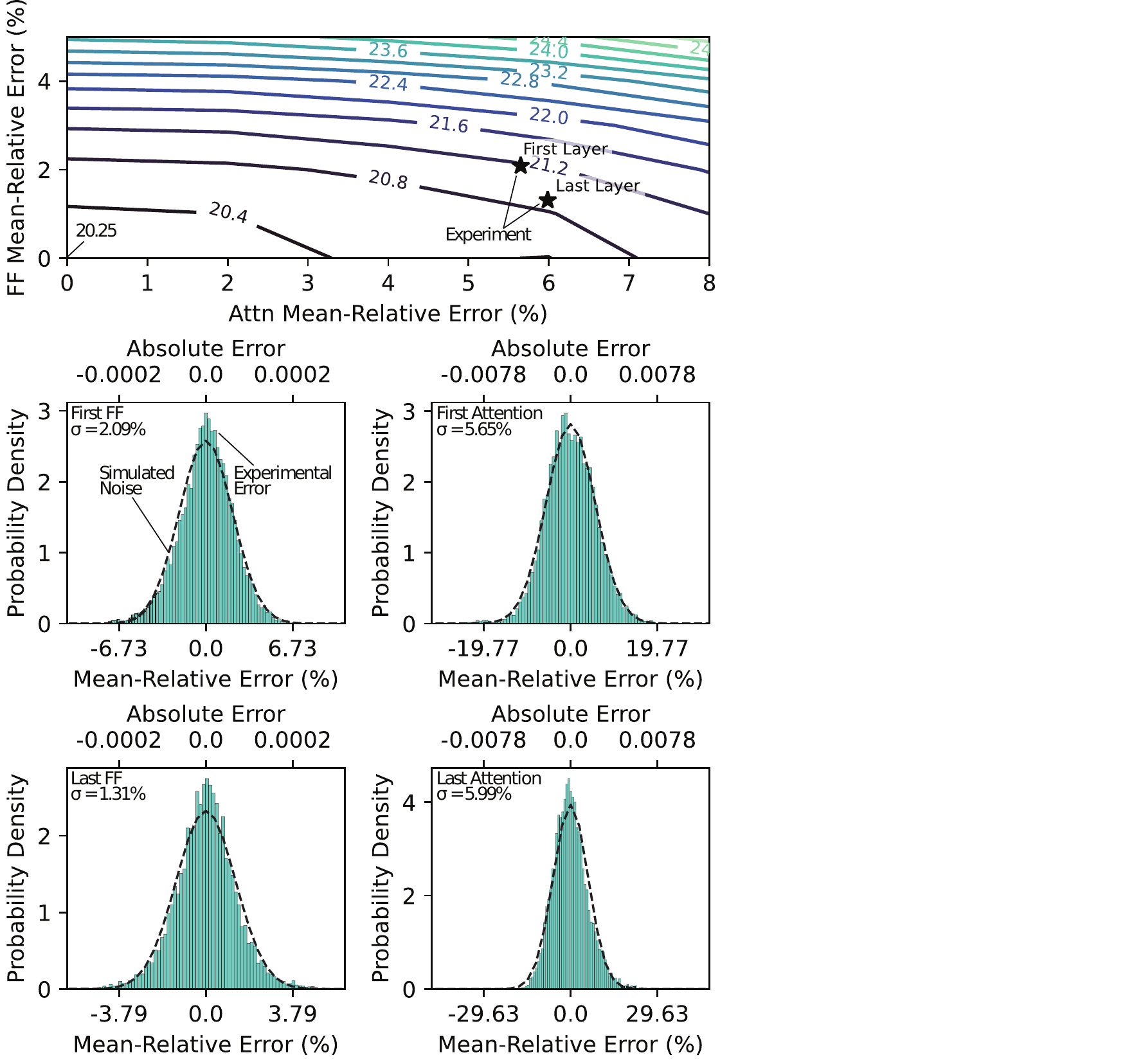}
    \caption{\textbf{Comparison of experimental noise and simulated Optical Transformer noise tolerance.} Top: Simulated performance (Wikitext-103 validation perplexity (PPL), shown as contours) versus percent mean-relative noise in feed-forward (FF) and attention (Attn) layers. Noise levels from experimental data marked with a star for dot products sampled from first and last Transformer encoder layers. Bottom: comparison of simulated noise model to error from experimental data. The Gaussian shape of the simulated noise models the experimental errors accurately.}
    \label{fig:noise}
\end{figure} 

\begin{figure*}[htbp!]
  \centering
  \includegraphics[width=0.97\textwidth]{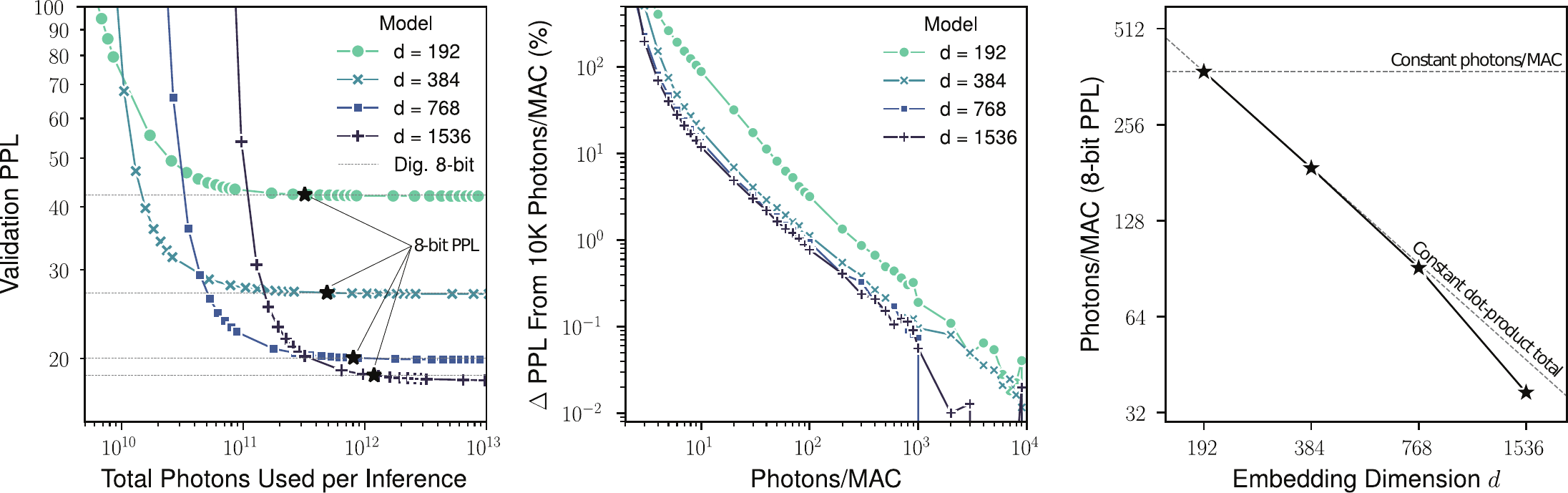}
  \caption{\textbf{Simulations of Optical Transformer behavior with varying photon usage.} Left: Wikitext-103 validation-set perplexity (PPL) versus embedding dimension $d$ and total photons used for a single inference (predicting next token in language modelling, or processing one sequence in a classification task). 8-bit digital model performance is shown with dashed lines. With sufficiently large numbers of photons, optical hardware can achieve the same perplexity as digital-electronic hardware, under the assumption that the optical hardware's precision is limited by photon shot noise. Middle: Percent change in perplexity from the perplexity achieved when using $10^4$ photons per multiply-accumulate (MAC), versus photons-per-MAC. At $10^4$ photons-per-MAC, the perplexity is approximately as good as one can achieve, so this plot shows how the perplexity degrades from ideal as one uses fewer photons-per-MAC; the plot exhibits truncated power-law scaling. Right: Scaling of number of photons needed for an Optical Transformer to achieve the same perplexity as an 8-bit digital-electronic processor, versus model size.  }
  \label{fig:photonresults}
\end{figure*}

\subsection{Error Tolerance and Simulation Accuracy (Experiments Using Optical Hardware and Comparison to Simulation)}

 We determined experimentally that Transformer operations are able to run on real hardware without severely degraded performance from systematic errors. In our experiments, as described in \cref{ss:expt}, we used high photon numbers and averaged results of multiple runs per dot product in order to eliminate noise, leaving only the hardware's systematic errors. In the bottom four panels of \cref{fig:noise} are the histograms of the experimental differences from correct values. The simulated noise distributions (dotted lines) match well with the experimental data, which confirms that they are an accurate representation of the real systematic error behavior. \cref{fig:noise} (top) is a map of the performance of the simulated model over different configurations of the mean-relative (in percent) noise at every layer of feed-forward and attention blocks. The model performs well with significant noise (experimental noise levels marked with stars), within 1 perplexity from noise-free performance unless the noise is very high. The model is more resilient to noise in the attention operations than in the linear layers. These results show that Transformers can run well when subject to the noise of a real accelerator system, and that our digital model of the system is a plausible approximation of how a real one might behave.

 While we found 8-bit precision to be stable for QAT, the optical Transformer can perform inference at lower precision, as implied by its error tolerance. To study this further we conducted a simple ablation on the input and output precisions used at inference, on the 8-bit-QAT base-sized model with LUT. The details of this study are in \cref{appendix:ablation}.

\subsection{Optical Scaling Laws (Simulation)}
\label{ss:scaling}
When systematic errors are not the limiting factor in performance, optical shot noise becomes relevant and limits a model's energy efficiency. To study how optical Transformers scale and behave with different optical resources available, we removed any systematic error in our simulations, used the same EMA-style QAT \cite{jacob2018quantization} (exponential moving averages maintained for min/max model statistics) for all models, and observed their behavior with different photon budgets and simulated shot noise, as described in \cref{ss:sim}.

Optical Transformers achieve language modelling performance close to their digital counterparts' when shot-noise-limited at modest photon budgets. The perplexities on the Wikitext-103 validation set of various optical Transformer models simulated with different total photon usage (amount used for input data) are shown in \cref{fig:photonresults} (left). The curves illustrate a tradeoff: larger models need larger photon totals to function well, and there are different optimal model choices based on the photon budget. We define photons/MAC as the total photon budget (amount at input) divided by total MACs. The percentage difference from the performance at 10K photons/MAC (\cref{fig:photonresults}, middle)---chosen to represent an ideal high-precision scenario---is roughly power-law scaled in photons/MAC for all models with truncation near 10K; better performance can be had with more photons, but with diminishing returns, and the performance matches or exceeds that of the 8-bit digital models' when the photon budget is not too low ($\sim 10^2$).

The models use fewer photons/MAC as they scale, achieving the theoretical efficient scaling where the total per-dot-product photons needed is constant. To study how photon usage scales, we determined how many photons it takes to reach the performance of 8-bit digital models. These values, in \cref{fig:photonresults} (right), decrease nearly as $\frac{1}{d}$---the total photons needed per dot product is constant (bottom dashed line). The Transformer architecture clearly takes advantage of efficient optical scaling with larger model sizes. In fact, smaller per-dot-product totals are required for the largest model, suggesting that larger Transformers may require less output precision. This is consistent with other work which found that precision requirements are constant or reduced with scale \cite{train_large_then_compress}. Meanwhile, the already low photon usage of the largest model suggests that models larger than our simulations (>10B parameters) may use <1 photon/MAC. This sub-photon operation works in optical systems \cite{Wang2022, netcast} and is in essence no different at all from operation at higher photon counts (since the number summed at detection is still high).

These empirical scaling results are tied to our specific configurations and training strategies. Depending on the scales and dynamic ranges of inputs and weights, different amounts of photons may be transmitted to the output; the statistics of a model affect its efficiency. In Appendix \ref{appendix:clipping} we explore a different scheme, but the effects of different methods remains an interesting topic for future work.

\subsection{Estimated Energy Usage (Simulation)}
\label{sec:energy_analysis}

\begin{figure*}[htb!]
  \centering
  \includegraphics[width=0.98\textwidth]{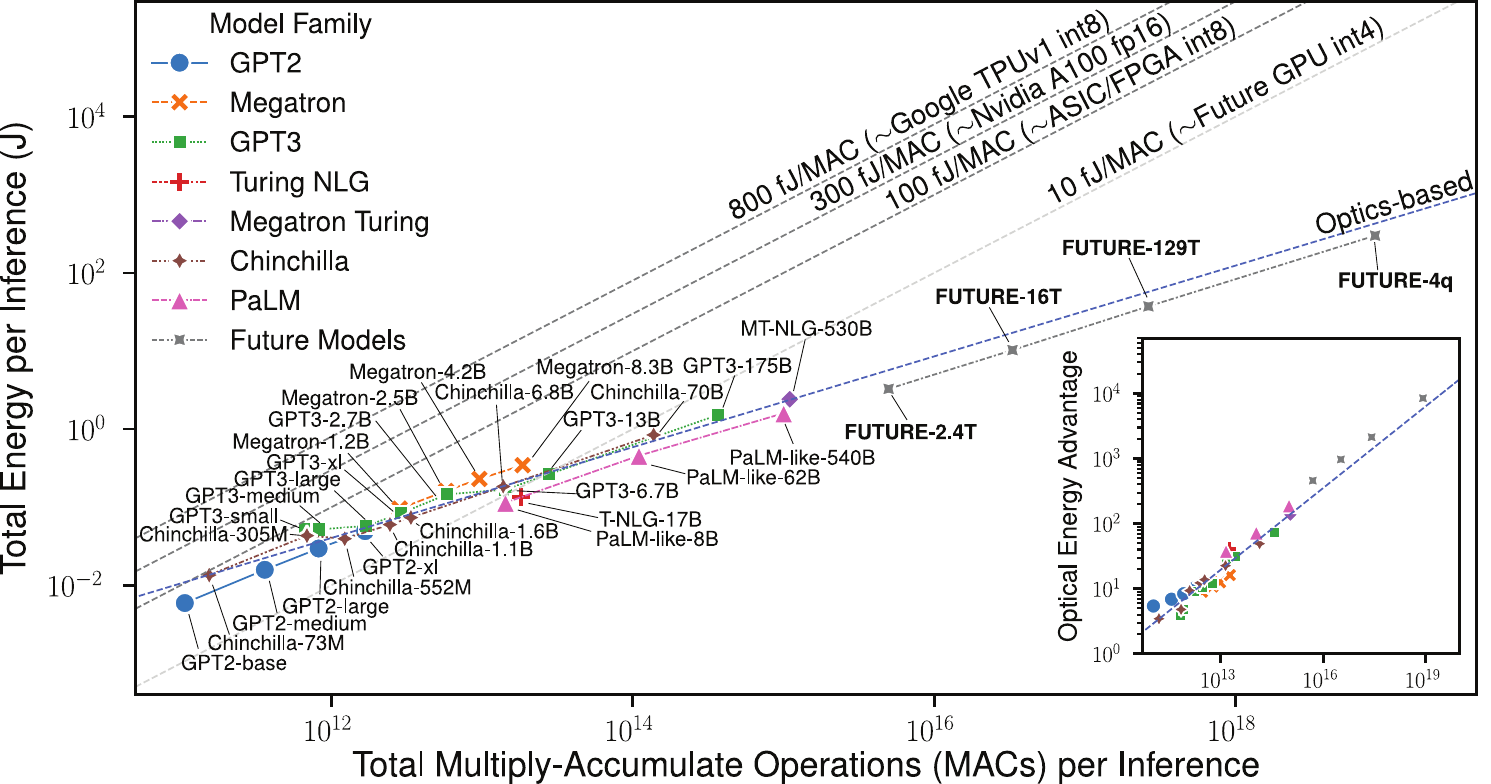}
  \caption{\textbf{Estimated energy usage of Transformer models on optical hardware for a single inference (predicting the next token in language modelling, or processing one sequence in a classification task).} Hypothetical future model designs are labelled \textbf{FUTURE-*}. Estimated energy/MAC for digital systems is based on \cite{reuthersurvey2020}. Trend for energy usage in optical systems (blue) computed based on real models only. Inset: energy advantage of running on optics over estimated NVIDIA A100 usage. The advantage grows with the model compute. $\mathrm{M}=10^6$, $\mathrm{G}=10^9$, $\mathrm{T}=10^{12}$, $\mathrm{q}=10^{15}$ parameters.}
  \label{fig:energy}
\end{figure*}

The efficient photon scaling trend we observed in \cref{ss:scaling} suggests that Transformers running on optical hardware could achieve significant energy efficiency advantages over running on digital hardware. To understand the efficiency of Transformers on optical hardware, we designed an ONN system based on current hardware that is like our experimental setup, with our measured precision and photon scaling. It is an inference system with in-place weights which are loaded once and reused forever, activations read from and written to SRAM for every layer, a 10 GHz light modulator array, and an optical ``core'' which can perform 10M multiplications per cycle (this can be thought of as a 10 megapixel SLM). The photon-per-MAC scaling versus model dimension is taken to be the $1/d$ scaling which we found was possible in our simulations, and we assumed that the model operates with 5-bit input precision, 8-bit weight precision, and 7-bit output precision, as determined by our study of low precision performance in \cref{appendix:ablation}.

We used this model to estimate the costs of running Transformers. We followed the approach in \cref{par:energy}. For electrical energy we calculated the cost of loading and detecting every value in the operation of the model. For attention, weights-in-place is not possible, as the matrix products are between activation data. In this case, a different scheme which can load both operands dynamically \cite{hamerly2019large} is desirable. In any case, this means that loading and encoding both operands must be counted towards the energy cost. In general, the electrical costs are composed of amplification, modulation, memory access, and digital/analog conversion. The calculation of optical energy is simple: the total number of MACs for a model was computed, and multiplied by the photons/MAC and energy of a photon. The photon usage was assumed to scale according to the constant-per-dot-product rule we verified in our simulations. In Appendix \ref{appendix:energy} we detail our method for estimating energy usage, and assumed quantities based on contemporary hardware.

As models grow, running Transformers on optical hardware has a large and asymptotic efficiency advantage over running on digital hardware. In \cref{fig:energy} we chart estimates of the forward pass energy required for various models\footnote{The recent PaLM \cite{palm} models used a modified architecture. For simpler comparison, we make our estimates using a model with GPT-like architecture but with the PaLM model dimensions, which we call PaLM-Like.}, including a hypothetical family of large, dense Transformer models designed in a similar fashion, which we label \textbf{FUTURE-*}. These represent the sizes of models in $\sim$ 10-15 years if current design trends hold. For comparison, we also chart various digital systems \cite{reuthersurvey2020} in different performance regimes, and a hypothetical ``next generation'' GPU that can use $\sim$10 fJ/MAC. For small models, the optics-based system uses about the same energy, but eventually gains an advantage that scales asymptotically with the number of MACs. For the larger models, MT-NLG-530B and FUTURE-4q, the optics-based approach would have $\sim$$140\times$ and $\sim$$8500\times$ energy advantages over the current state-of-the-art GPU (NVIDIA A100) respectively.

\begin{figure*}[htb!]
    \centering
    \includegraphics[width=0.97\textwidth]{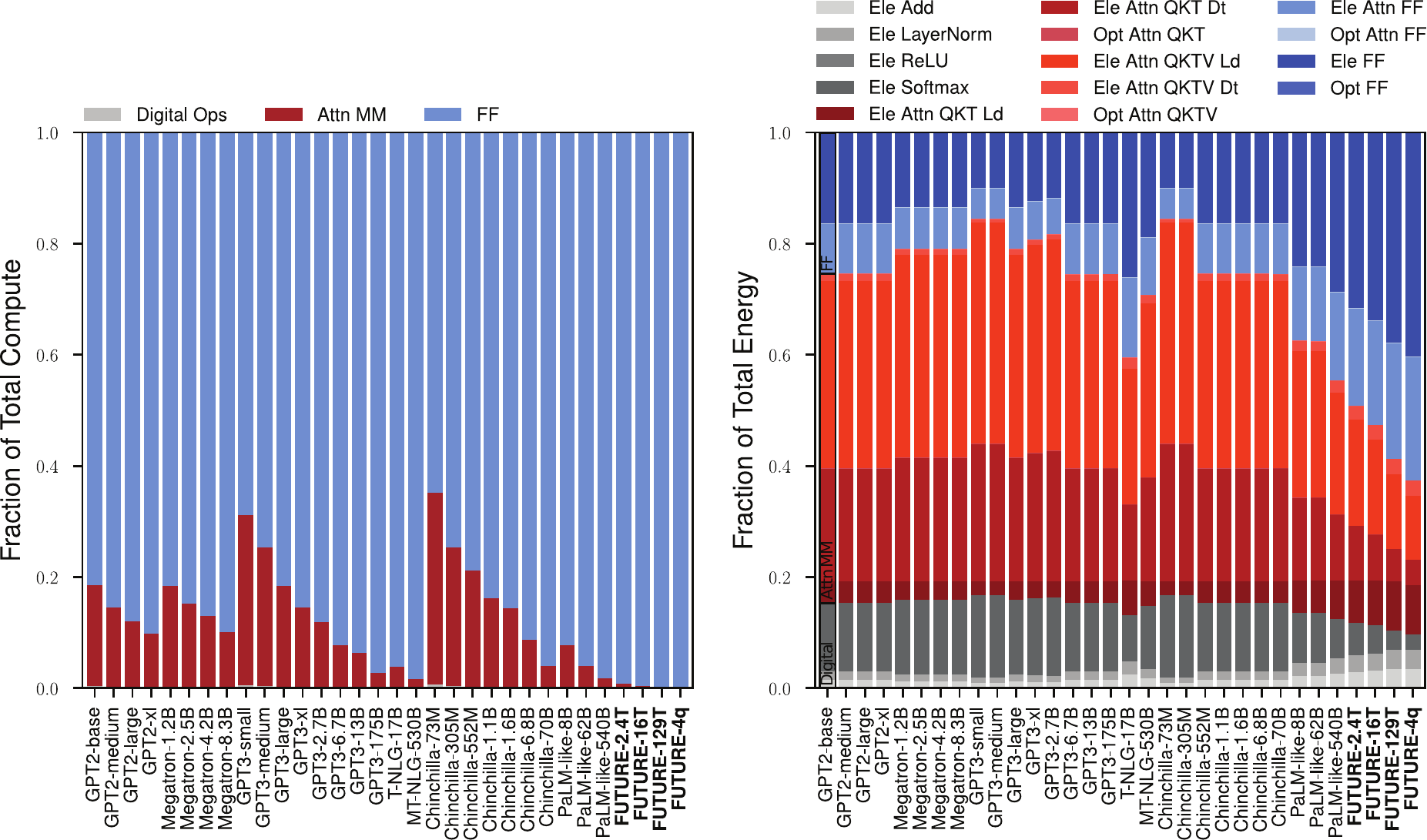}
    \caption{\textbf{Breakdown of energy costs for Optical Transformers.} Left: fraction of total compute used by digital operations, attention (Attn) matrix-matrix multiplications (MM), and feed-forward (FF) components. Feed-forward layers account for most of the compute. Right: breakdown-of-costs for models by layer. The energy costs of attention operations is expensive. ``Ele *'' operations: electrical costs of loading (Ld), detecting (Dt), or both for data for the operation. Operations related to attention computation (ie. $QK^T$ and $\mathrm{softmax}(\frac{QK^T}{\sqrt{d_h}})V$, where $d_h$ is the embedding dimension for an attention head) are expensive for little compute. Functions computed digitally have their energy costs estimated as the cost of reading and writing to memory the required data. $\mathrm{M}=10^6$, $\mathrm{G}=10^9$, $\mathrm{T}=10^{12}$, $\mathrm{q}=10^{15}$ parameters.}
    \label{fig:energy_breakdown}
\end{figure*}

Breaking down the sources of compute and energy costs in Transformer models running optically illustrates how aspects of model/system design affect energy usage. The breakdown of compute and energy costs by source is in \cref{fig:energy_breakdown}. We find that as models get larger the feed-forward layers require most of the computation, but that the energy of data access in attention is still very expensive. This is because of the need to save/load many attention matrices from memory, and the fact that a weights-in-place scheme cannot be used for the matrix-matrix products because the products are between activations. In total, this means that attention layers have high energy costs for small amounts of computation. Existing model design trends have moved towards focusing much harder on feed-forward layers, and so for the largest real (and our hypothetical future) models the fraction of energy cost taken by attention is low. Finally, we note that the operations we assume run on digital computers---such as nonlinear functions, in gray---do not account for much of the total energy cost (though they too are a small fraction of the total compute). 

As a result the energy scaling of Transformer models is better than what only the favorable photon scaling and free data transport would predict; these trends in the design of real models have increasingly favored optics over time. Specifically, attention loads/stores a $n \times n$ attention matrix for each of the $h$ attention heads. Since attention heads perform few computations per data access (and MLP layers perform many), models with more MLP compute per attention head have a larger overall ratio of computation to energy usage; larger $\frac{d}{h}$ is more efficient. And real models are scaling $\frac{d}{h}$ over time: The largest GPT2 \cite{gpt2} uses $\frac{d}{h} = 64$, while the larger GPT3 \cite{gpt3} variants use $\frac{d}{h} = 128$. MT-NLG-530b \cite{MT-NLG} uses $\frac{d}{h} = 160$. Finally, the recent PaLM \cite{palm} models use significantly larger $\frac{d}{h}$, up to $384$.

We further analyze what components of the system affect the energy costs in \cref{appendix:energy_breakdown}.

\section{Discussion}
The results given in \cref{sec:energy_analysis} on optical Transformers' efficiency have implications for the design of future hardware and software systems that might run them.

\subsection{Implications for Designing Optical-Neural-Network (ONN) Accelerators}

A functioning ONN system requires computation units (which we call \textit{cores}) for performing MACs, detectors, modulators, and memory. We define a \textit{core} as any component that can perform element-wise products with locally stored/maintained weights in parallel. For example, a 10-megapixel spatial light modulator (SLM)---which can perform $10^7$ scalar multiplications at the same time, if one weight is stored per pixel---could be considered a core. The biggest challenge in creating ONN hardware is in having enough (or large enough) cores, as the number required to run a layer in a single pass scales quadratically with $d$ (since they must maintain all the weights).  In \cref{appendix:chunking} we examine what specifications might be required to build such a large system. To deal with this enormous hardware requirement, it is possible to introduce new hardware-time-energy tradeoffs. One strategy for doing this is to split up the large weight matrix into several ``chunks'', and cycle through the chunks on the available cores, computing the MVMs, and collecting the results. This comes at the cost of latency, and requires energy to frequently reload inputs to be reused for more weights. In \cref{appendix:chunking} we further explore the consequences of this tradeoff (compare to GPU in the similar case where multiple processors are required). Based on these findings and our energy calculations, we identify areas for future efforts in ONN hardware design:
\begin{itemize}
\item Once matrix-matrix product operands exceed $10^4 \times 10^4$ in size the advantage is significant, and therefore a future ONN should implement at least this level of parallelism to achieve $>$$100\times$ efficiency improvements over current state-of-the-art GPUs (NVIDIA A100). Systems with many cores could run larger models even more efficiently. For example, a system with $\sim$3000 cores with $10^7$ weights each could enable dense Transformer models with over 10 trillion parameters to run $\sim$1000x more efficiently than on a current state-of-the-art electronic processor with an efficiency of $\sim$300 fJ/MAC (such as an NVIDIA A100 GPU). 
\item Given the assumptions we made about weight-maintenance costs in making our estimates (5.6 \textmu W per weight; see \cref{appendix:energy}), an Optical Transformer would need to operate in the regime where a single matrix-vector multiplication is performed every 0.1 nanoseconds (about 20 TFLOPs of throughput). Arrangements with larger or multiple cores could achieve very large throughput easily, and without data transport  energy costs; A system with a hundred large ($\sim$$10^8$ weights each) fully-utilized cores could, for example, reach $\sim$10--100 EFLOPs of throughput.
\item Compared to our assumptions (\cref{appendix:energy}, based on existing component energy costs and \cref{appendix:chunking} for hardware scale requirements), current ONN prototypes either operate at much lower clock rates or at much smaller scale. Thus while we have shown that a system using existing components may achieve our projected $100\times$ advantage, building a full ONN system that is simultaneously fast, large-scale, and efficient is still an open challenge.
\item Many of the drawbacks and limitations of ONNs could be eliminated if future work finds an easy way to implement optical ``memory'' for re-reading data.
\item Weights-in-place (\cref{ss:wip}) is a good choice for large-scale neural network designs that are popular today, since their foundation is large linear operations repeated across many tokens/batches.
\item Because data access costs are the limiting factor in ONN efficiency, future improvements in CMOS technology (DAC, ADC, memory access costs in particular) will be greatly beneficial. For example, our future projection of halved DAC/ADC costs and 4-bit inference would roughly double all of our energy advantage estimates. In \cref{appendix:future_energy} we combine this with the assumption that future memory access may be 5$\times$ cheaper to show that future optics-based systems might achieve energy advantages of $>$$100,000\times$ running models the size of FUTURE-4q (over our 300-fJ/MAC state-of-the-art GPU estimate).
\end{itemize}

Note that we are determining the future advantage against current hardware so that the projections can be compared to our estimates in \cref{sec:energy_analysis}; future electronics improvements would of course also advance the performance of digital systems, but it is clear at least that ONNs benefit from electronics improvements too, even if future GPUs might have advanced significantly from what exists today. Furthermore, since ONN accelerators' (running Transformers) energy consumption is never compute-dominated, they stand to benefit at least as much as digital systems from improvements to electronics for efficient data access. 

\subsection{Scalable Optical Deep Learning}
Much like how Transformer designs have become increasingly favorable for optics, new approaches for managing large models might solve scalability problems for ONNs. For example, sparse-expert architectures \cite{switch_xf, sparse_expert_design} shard computations to different parts of the system, avoiding large input communication/reloading costs. Another example is the myriad research investigating how to run large models, including Transformers, at lower precision \cite{q8bert, vit_quant, xf_quant, llm_int8}. This directly benefits ONNs as their energy costs are driven by data access, and cannot operate with high precision. So, while there are speed/energy tradeoffs and communication cost concerns in ONNs, research is already exploring solutions for the analogous problems in digital systems.

A prospect for future work, though, is designing and scaling models for optimal efficiency on optical hardware. Presently, there is plenty of work in exploring compute/time/cost tradeoffs on digital hardware, including derivation of sets of scaling laws for how to optimally grow models \cite{Tan2019EfficientNetRM, Zhai_2022_CVPR_scaling_vits, kaplan_scaling_laws}. For optics-based systems, energy costs per FLOP and per memory access are different and model statistics affect energy usage. Therefore, future work could focus on performing the same kinds of scaling and tradeoff analysis in the context of the new hardware, deriving trends and design patterns for future neural network architectures. We expect that these new trends might result in different recommendations for scaling width over depth, the use of activation functions which shape activation and weight distributions, and so on. For example, if a model is scaled with width, the compute scales quadratically with the amount of data accesses, but only linearly with depth; a single-encoder-layer GPT3 (ie. a model with the same parameter count) running optically could be roughly $20\times$ more efficient; a $\sim$$1600 \times$ advantage over GPU instead of $\sim$$80\times$ in our estimations. The tradeoffs between the energy benefits of designing models this way and how well they perform on various tasks is also worth future investigation.

Our results and insights on designing deep-learning models for, and running them on, optical hardware can be summarized as follows:

\begin{itemize}
\item For executing Transformers, optical hardware has an asymptotic efficiency advantage over digital electronics as model sizes increase - we have shown that Transformers executed optically have a per-MAC energy cost that scales as $\frac{1}{d}$ (or better) where $d$ is the width of the model. This is not possible in digital systems, where energy per computation is typically constant.
\item The largest models today are already large enough to benefit significantly ($>$$100\times$ estimated efficiency gain running optically over NVIDIA A100 GPU). We project that optical accelerators could have a $>$8000$\times$ energy advantage when running future, larger models.
\item Many modern deep-learning strategies and architectures (such as Transformers) have implicitly or explicitly been designed to perform well on GPU-based systems by enabling greater parallelism and more efficient use of memory. In the past, models designs were such that operations could not easily be run in parallel, but Transformers sought to make large models run efficiently by exploiting hardware's strengths in perform large, parallel, dense calculations, and improved in this aspect as they scaled. As a consequence, as Transformers (and other deep learning models) continue to be optimized for parallel digital electronic hardware, they will continue to become even more efficient on optical hardware.
\item Models that perform well at lower precision are maximally beneficial---they are easier to run on noisy, analog optical hardware, and minimize data access costs.
\item Architectures that perform more computations per data access will be most energy-efficient. Recent models with a stronger focus on linear operations \cite{tolstikhin2021mixer, gmlp} could be promising for optical implementation. More generally, an important direction for future work is to design neural-network architectures that are optimal for optical matrix-vector multipliers, for example by maximizing data reuse.
\end{itemize}

\section{Summary and Conclusion}
We have demonstrated the ability of Transformer models to run accurately and efficiently on optical hardware through optical experiments and numerical simulations. We examined Transformers' scaling behavior with optics and used our findings to show that optical systems have a large and asymptotic energy advantage over digital ones that \textit{grows} with the model size. For example, we showed that optical hardware may achieve an over $100\times$ energy advantage when running the largest Transformer models today ($\sim$500 billion parameters) and that larger, future Transformers ($\sim$4 quadrillion parameters) may be realized with an $>$$8000\times$ optical energy advantage. We then explained how existing trends in the machine-learning community (such as optimizing Transformer designs for better parallelism, and building models that rely more heavily on multi-layer perceptrons within them \cite{tolstikhin2021mixer}) are resulting in models that are increasingly better suited for optical hardware than the canonical Transformer model \cite{transformer} and previous language models---they involve large, dense, linear calculations to saturate processors' compute capabilities without data-access bottlenecks. This is not a coincidence; in general, neural-network models that are designed to be more efficient on digital-electronic hardware by maximizing compute and parallelism per-data-access will be more efficient on optical hardware, where data-access costs are also a primary concern.

We discussed what specifications optical hardware would need to meet to realize the projected energy-efficiency advantages for different model sizes. For example, to realize the 100$\times$ energy-efficiency advantage for current Transformer models, we project needing optical matrix-vector multipliers each capable of multiplying a $10^4 \times 10^4$-dimensional matrix with a $10^4$-dimensional vector at a rate of 1 matrix-vector multiplication every 0.1 nanoseconds, with power consumption of roughly $\sim$1 pJ/bit for input/output data access and conversion. While existing components are capable of this level of energy-efficiency, there is an open hardware systems-level challenge of making an optical neural network that is simultaneously big, fast, and efficient enough. We also discussed how future architectural changes and improvements to electronics would further improve ONN efficiency---with better model quantization and assumptions about near-future hardware, future Transformers running on optical hardware could exceed a 100,000$\times$ energy advantage over current state-of-the-art $\sim$300-fJ/MAC digital electronics. The efficiency of digital-electronic neural-network accelerators is improving at a rate of $\sim$10$\times$ every 7 years \cite{digital_energy_trends}, suggesting that even if the proposed optical hardware takes 10 years to develop, there should still be a several-orders-of-magnitude benefit to using optics for neural-network computations when the scaled hardware becomes available. Furthermore, any improvements to the energy efficiency of electronic memories directly benefits both digital processors and optical processors, so future development of digital processors may only shrink the optics--electronics efficiency gap through improved MAC-computation efficiency, which has proceeded predictably for the past two decades based on transistor scaling \cite{transistor_scaling}.

We believe our findings about the potential energy-efficiency of optical accelerator hardware strongly motivate pursuing the development of optical processors for large-scale deep learning with Transformers or other models that heavily rely on weight-stationary matrix-vector multiplication.

\section{Acknowledgements}
The authors wish to thank NTT Research for their financial and technical support. Portions of this work were supported by the National Science Foundation (awards CCF-1918549 and CBET-2123862) and a David and Lucile Packard Foundation Fellowship. We acknowledge helpful discussions with and feedback from Alen Senanian, Benjamin Malia, Fan Wu, Federico Presutti, Jeremie Laydevant, Sridhar Prabhu, and Vladimir Kremenetski.

\section{Author Contributions}
M.G.A., T.W., L.G.W., and P.L.M. conceived the project. M.G.A., S.M., T.W., L.G.W., and P.L.M. designed the experiments. S.M. performed the dot-product computation experiments on the SLM-based ONN setup. M.G.A. and S.M. analyzed the experimental data. M.G.A. designed and trained the optical Transformer models, and simulated the effects of optical hardware running them. M.G.A. and T.W. created the model of an ONN accelerator for estimating ONN energy consumption for running Transformers. M.G.A., S.M., T.W., L.G.W., and P.L.M. wrote the manuscript. P.L.M. supervised the project.

\section{Reproducibility Statement}
We provide code, data, and instructions to reproduce our results. 

Our models, configurations, procedures, and assumptions for theoretical calculations are provided in the appendices. Configurations of optical Transformer models, training hyperparameters, and quantization are available in \cref{appendix:training}. A description of our experimental procedure including the techniques and components used is in \cref{appendix:expt}. The assumptions we used for our calculation of estimated ONN energy costs are detailed in \cref{appendix:energy}. The assumptions about ONNs, electronic hardware, and model precision for future systems are in \cref{appendix:future_energy}.

We make our code and data available so that all of our analysis can be reproduced. We provide data collected from our experiments, code to estimate energy consumption of ONNs running Transformers, and the source code for our training/evaluation/simulations of optical Transformer models. Finally, the release contains code to analyze all of the data and results. Instructions for how to perform each of these tasks is provided in the README file. All of these materials are available at: \url{https://doi.org/10.5281/zenodo.7574198}.

\clearpage

\bibliography{citations}
\bibliographystyle{icml2023}

\clearpage

\section*{Appendices}

\begin{appendix}
\renewcommand{\bottomfraction}{1.0}
\renewcommand{\topfraction}{1.0}
\renewcommand{\textfraction}{0.1}

\section{Optical Transformer Training Hyperparameters}
\label{appendix:training}

\begin{table}[]
    \caption{Model configurations for optical Transformers. $\mathrm{M}=10^6$.}
    \centering
    \begin{tabular}{l|llll|l}
        \toprule
        Model & n & d & h & L & Non-emb. Params  \\
        \midrule
        Tiny & 1024 & 192 & 12 & 12 & 15M \\
        Small & 1024 & 384 & 12 & 12 & 40.6M \\
        Base & 1024 & 768 & 12 & 12 & 123.7M \\
        Large & 1024 & 1536 & 12 & 12 & 416.3M \\
        \bottomrule
    \end{tabular}
    \label{tab:opt_xf_models}
\end{table}

\begin{table*}[t]
    \caption{Pretraining hyperparameters for optical Transformer models. All models were trained with the \textbf{AdamW} \cite{adamw} optimizer.}
    \centering
    \begin{tabular}{c|ccccccccccc}
        \toprule
         Model & Steps & Batch & lr & $\beta_1$ & $\beta_2$ & $\epsilon$ & Weight decay & Dropout &  Schedule & Warmup & Stop \\
         \midrule
         Tiny & 90000 & 32 & 2e-4 & 0.9 & 0.999 & 1e-8 & 0.02 & 0.1 & Cosine & 2500 & - \\
         Small & 90000 & 32 & 2e-4 & 0.9 & 0.999 & 1e-8 & 0.02 & 0.1 & Cosine & 2500 & - \\
         Base & 90000 & 32 & 2e-4 & 0.9 & 0.999 & 1e-8 & 0.02 & 0.1 & Cosine & 2500 & - \\
         Large & 90000 & 32 & 2e-4 & 0.9 & 0.999 & 1e-8 & 0.02 & 0.1 & Cosine & 2500 & 82500 \\
         \bottomrule
    \end{tabular}
    \label{tab:pretraining}
\end{table*}

\begin{table*}[t]
    \caption{Quantization aware training hyperparameters for optical Transformer models. All models were trained with the \textbf{RMSProp} \cite{rmsprop} optimizer. Quantization parameters are in Table. \ref{tab:quantization}.}
    \centering
    \begin{tabular}{c|cccccccccc}
        \toprule
         Model & Steps & Batch & lr & $\alpha$ & $\epsilon$ & Weight decay & Dropout & Schedule & Warmup & Stop \\
         \midrule
         Tiny & 7327 & 64 & 1e-5 & 0.99 & 1e-8 & 1e-5 & 0.1 & Cosine & 2500 & - \\
         Small & 7327 & 64 & 1e-5 & 0.99 & 1e-8 & 1e-5 & 0.1 & Cosine & 2500 & - \\
         Base & 7327 & 64 & 1e-5 & 0.99 & 1e-8 & 1e-5 & 0.1 & Cosine & 2500 & 5500 \\
         Large & 7327 & 32 & 1e-5 & 0.99 & 1e-8 & 1e-5 & 0.1 & Cosine & 2500 & 5500 \\
         \bottomrule
    \end{tabular}
    \label{tab:qat}
\end{table*}

\begin{table*}[t]
    \caption{Hyperparameters for optical Transformer Quantization. We perform QAT with both a percentile-clipping approach and by clamping based on an exponential moving average (EMA) of model statistics with factor $\gamma$. For the Base-sized model that is used in our experiments (LUT-Base), we use lookup tables (LUT) for inputs and weights instead of quantization.}
    \centering
    \begin{tabular}{ccc|c|ccc|ccc}
        \toprule
        \multicolumn{3}{c|}{Overall Config} & \multicolumn{1}{c|}{EMA} & \multicolumn{3}{c|}{Attention Clipping} & \multicolumn{3}{c}{Feed-Forward Clipping} \\
        Model & Precision & Rounding & $\gamma$ & $\mathrm{Input_1}$ & $\mathrm{Input_2}$ & Output & Input & Weights & Output \\
        \midrule
        Tiny & 8-bit & Stochastic & - & 99.99\% & 99.9\% & 99.9999\% & 99.99\% & 99.9\% & 99.9999\% \\
        Small & 8-bit & Stochastic & - & 99.99\% & 99.9\% & 99.9999\% & 99.99\% & 99.9\% & 99.9999\% \\
        Base & 8-bit & Stochastic & - & 99.99\% & 99.9\% & 99.9999\% & 99.99\% & 99.9\% & 99.9999\% \\
        Large & 8-bit & Stochastic & - & 99.99\% & 99.9\% & 99.9999\% & 99.99\% & 99.9\% & 99.9999\% \\
        LUT-Base & LUT & Stochastic & - & 99.99\% & 98\% & 99.9999\% & 99.99\% & 99\% & 99.9999\% \\
        Tiny & 8-bit & Deterministic & 0.999 & - & - & - & - & - & - \\
        Small & 8-bit & Deterministic & 0.999 &  - & - & - & - & - & - \\
        Base & 8-bit & Deterministic & 0.999 & - & - & - & - & - & - \\
        Large & 8-bit & Deterministic & 0.999 & - & - & - & - & - & - \\
        \bottomrule
    \end{tabular}
    \label{tab:quantization}
\end{table*}

The optical Transformer models were pretrained on the Wikitext-103 \cite{wikitext_103} dataset and used the same tokenizer as GPT2 \cite{gpt2}. All models used \textbf{Xavier uniform initialization} \cite{xavier_init}. The architectures are in \cref{tab:opt_xf_models}. Embedding layers were initialized with a normal distribution with $\sigma = 0.02$. We used the AdamW \cite{adamw} optimizer, with weight decay applied to parameters which were not embedding, gains, or biases. Dropout was applied after every linear layer (including those in attention), as well as on the attention matrix and after the $\textrm{softmax}({\frac{QK^T}{\sqrt{d_h}}})V$ product in the attention calculation. The values of the parameters used for the training scheme are in Table \ref{tab:pretraining}.

After pretraining the models were quantized via our 8-bit QAT scheme. For QAT we used the RMSProp optimizer \cite{rmsprop}. The parameters we used for the training are in Table \ref{tab:qat}. To clamp weights and activations we employ two different approaches: first,  we kept running statistics of minimum and maximum values with an exponential moving average (EMA, with parameter $\alpha$) for every layer and use those to clamp. Second, we recorded the minimum/maximum statistic throughout the network for a forward pass to apply a clipping scheme. Specifically, we clamped weights and activations to percentiles of the maximum values collected for each layer. The outputs were either rounded to the nearest integer during QAT, or stochastically rounded to nearby values. Finally, for the Base-sized model we used to run the experiments, we directly used the lookup tables (LUT) instead of ``simulating'' the quantization of inputs and weights (though outputs are still quantized). Table \ref{tab:quantization} details our use of these various techniques in the models.

For evaluation we used the perplexity (PPL) metric to measure the language modelling performance on Wikitext-103. We evaluated the perplexity over the entire validation set, and ran the model with context length 1024 (the same as in training) and a 1024-token stride length.

\section{ONN Experimental Procedure}
\label{appendix:expt}
\subsection{Experimental Setup}
\begin{figure*}[h]
    \centering
    \includegraphics[width=0.98\textwidth]{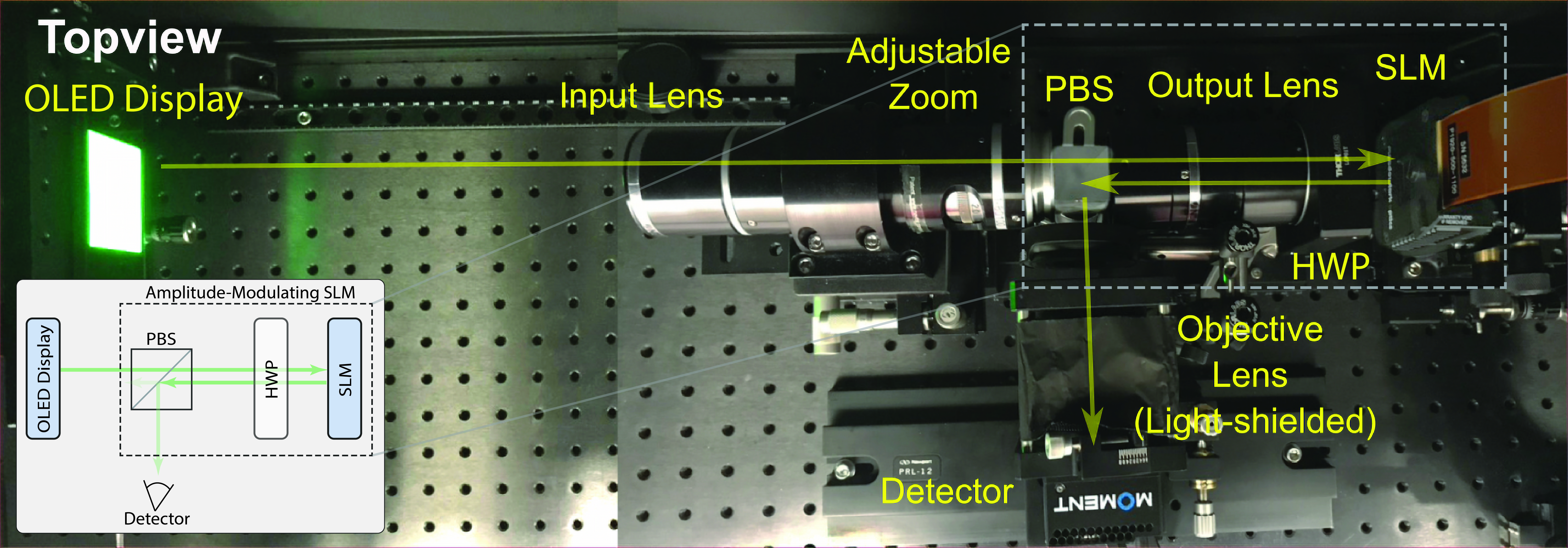}
    \caption{Photo of experimental setup used for running Transformer dot-product operations. Inset: simplified illustration of the experimental system. Spatial light modulator (SLM) + half-wave plate (HWP) + polarizing beam splitter (PBS) arrangement is effectively an amplitude-modulating SLM. The system works as follows: in our experiments, a vector is loaded as pixels on the organic light-emitting diode (OLED) display, and weights on the SLM. The input light enters through the PBS towards the SLM, passing through the HWP twice as the SLM reflects it. The SLM and HWP together rotate the polarization of the light, such that the amount reflected by the PBS towards the detector for each pixel is roughly the product between the pixel value and the corresponding weight on the SLM. The summation of these element-wise products by the detector yields the dot product. The figure is adapted from \cite{Wang2022} with permission.}
    \label{fig:apparatus}
\end{figure*}

\begin{figure*}[h]
    \centering
    \includegraphics[width=0.98\textwidth]{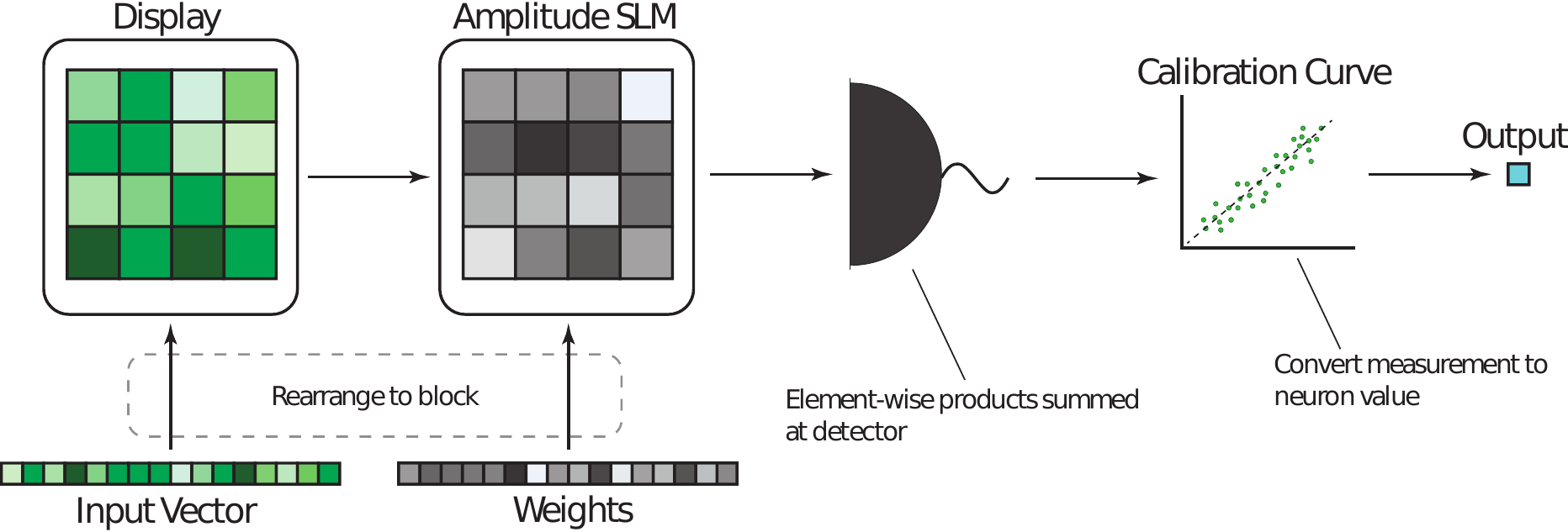}
    \caption{Simplified illustration of experimental setup operation. Weights are loaded and rearranged into a block on spatial light modulator (SLM) to prevent crosstalk between pixels of drastically different values. Data is rearranged on display accordingly. Measurements are looked up against calibration curve to obtain the final output value.}
    \label{fig:expt_illus}
\end{figure*}

Our setup is a SLM-based matrix-vector/vector-vector multiplier. The setup is shown in \cref{fig:apparatus} with a simplified illustration in \cref{fig:expt_illus}, and works as follows: Vectors corresponding to the inputs and weights are rearranged into squares of pixels and loaded onto the display and SLM respectively. They are aligned such that the light from display pixels will reach the corresponding pixels on the SLM. First, light from the display enters into the polarizing beam splitter (PBS), and reaches the SLM through a half-wave plate (HWP) which rotates its polarization. The phase is then modified by the SLM and reflected back through the half-wave plate, rotating the polarization again based on the phase difference. Then, the PBS only admits light of a certain polarization along one of its arms, aimed at a camera for detection. Summation of the output pixels is performed digitally. This SLM--HWP--PBS arrangement effectively creates an amplitude modulating SLM, where the output at each pixel is the element-wise product of the input pixel and corresponding weight pixel.

The OLED display has multiple color channels and a broad spectrum. For easier modulation by the SLM, we used a band-pass filter and only green light.

The components we used are:
\begin{itemize}
    \item Organic light-emitting diode (OLED) display (Google Pixel 2016)
    \item Reflective liquid-crystal modulator (1920-500-1100-HDMI, Meadowlark Optics)
    \item Half-wave plate (PH10ME-532, Thorlabs)
    \item Polarizing beam splitter (CCM1-PBS251, Thorlabs)
    \item Zoom lens for imaging onto SLM (Resolv4K, Navitar)
    \item Zoom lens and objective lens for imaging onto detector (1-81102, Navita and XLFLUOR4x/340, Olympus)
    \item Band-pass filter (FF01-525/15-25, Semroc)
    \item Camera for detection (Prime 95B Scientific CMOS Camera, Teledyne Photometrics)
\end{itemize}

This setup works as a good bench for testing the precision of optical Transformers by performing optical dot products involved in attention and feed-forward layers. Even though the optical dot products were performed one at a time, it is sufficient for showing that Transformer operations can run with the accuracy of ONNs, since matrix-vector and matrix-matrix products are merely collections of many dot products run in parallel.

\subsection{Calibration and Lookup Tables}
We used several techniques to reduce errors, map inputs to SLM/display values, and to convert detected outputs back to neural network values.

First, we developed a specialized data-pixel encoding scheme to reduce systematic errors. We noticed that a large source of error was with a limitation of our hardware---in particular the SLM pixels have cross-talk (pixels may affect their neighbors if they have very different values) and misalignment in the experimental setup may lead to corrupted outputs. To help with these issues, we created ``macropixels''---each input element (and weight) does not occupy one pixel on the display (SLM) but rather is mapped to a 3x3 grid of pixels, all with the same value. For the attention layers, we used 5x5 macropixels for the results we report, but later discovered that with 3x3 the performance is essentially the same. We also rearranged vectors into square blocks of pixels so that significantly nonzero weights are nearby each other \cref{fig:expt_illus}. For the vectors to better fit in the center of the field of view (where there is less distortion/misalignment) we computed the dot products using only the 400 largest weight elements (the corresponding input elements are loaded). While this may introduce some inaccuracy in the final results, we found that the benefits of computing the element-wise products more accurately outweigh the drawbacks of pruning the weights; the outputs were still quite accurate to the ground-truth dot-product values (\cref{fig:noise}). We suspect that this was the case because:
\begin{itemize}
    \item Transformer weights are not entirely dense; some weights were already zero.
    \item Because our setup only supports non-negative data anyway, we use the four-pass approach (\cref{sec:sim}). This means that for any given dot product, roughly half the weights and activations will be zero before considering the previously mentioned sparsity. 
    \item Meanwhile, a second consequence of this four-pass approach is that roughly half of activations will be zero as well, possibly rendering some of the pruned weights irrelevant.
    \item Transformers still perform well when pruned, and luckily larger models can be pruned more heavily \cite{train_large_then_compress}. While our pruning method is quite basic, the number of weights pruned was light (ie. $<75\%$) compared to what is possible with more advanced methods.
\end{itemize}
This approach was not necessary for attention calculations, since the dot products were sufficiently small to fit them entirely (64 elements).

Next, we consider the lookup tables (LUTs) of the display and SLM in the setup. In order to optimize the experimental results, the model used for experiment was trained to be aware of the realistic, discrete mappable values supported by the system. The display has a LUT with 256 unique levels (1000 levels total, but many are the same as others) and the SLM has roughly 128 unique levels (256 total). So they are roughly capable of 7 and 8 bit precision. The SLM also cannot fully extinguish input light---the minimum modulation is 2\% of the maximum transmission. Thus, the minimum absolute values of the weights were mapped to 0.02 instead of 0.

After applying these approaches, we finally collected the calibration curve, which maps the output intensity measurements to neuron values in the neural network and allows us to determine the experimental setup's systematic error. To do this we sampled randomly inputs and outputs of the layers we wished to run, computed their dot products both digitally and in experiment, and created a data set of experimental measurements and ground-truth-digital dot-product outputs. We then performed linear regression to find a mapping between experimental output and the correct values, effectively creating another lookup table. Then when future dot products were computed experimentally, the output was passed to this linear regression model (or it can literally be stored as a lookup table) to get the output. We used many photons and averaged outputs across multiple shots for each input, eliminating shot noise---any remaining error in this calibration scheme we defined as the system's systematic error.

It is important to note that in general other optical systems might have different causes of error from ours, but the overall accuracy of our system is representative of a typical ONN nowadays. 

\subsection{Model Design Optimization}

Transformers tend to have large dynamic ranges in their activations and weights \cite{xf_quant}. In particular, we found that systematic error is proportional to some characteristic amplitude of the output. So, because it scales roughly with the sizes of outputs, having large outlier values can increase the systematic error and worsen the calibration for all other values in the representable range. Furthermore, after quantization in a naive, linear scheme, large outliers mean that huge ranges of outputs which are seldom used are assigned to many of the quantization levels, while the rest of the small, common outputs are squashed into few buckets---so the model precision is poor. This can be an issue when quantizing any deep learning model, but was exacerbated here by those systematic errors and the fact that the lowest levels of the weights are 0.02 and not 0.0. Therefore, we opted for an aggressive clipping scheme and the clamped activation ReLU6 when training the model to be run (Appendix \ref{appendix:training}, \textbf{LUT-base} model); they reduce the dynamic range of inputs and weights and we found that they drastically improved the ONN's ability to run Transformer operations with smaller error. Having fewer values in the 0.02 bucket of the SLM LUT also improved QAT training stability significantly. Even though the non-zero light extinction at 0.02 is caused by the specific SLM in our setup, such issues may happen with other optical implementations made of elements with finite extinction or resolution, and here we described a method to mitigate such issues by modifying training methods.

\subsection{Transformer Dot Product Samples}
While the speed and parallelism limitation of our setup made it intractable to run an entire Transformer model on it, we attempted to sample dot products to run that were representative of the range of possible activation/weight statistics in the model. That way, our results would be very representative of what running the full model would be like. In particular, we found two ways in which statistics throughout the model vary: the statistics change with depth (shallow and deep layers behave differently) and operation type (matrix-matrix multiplication in attention has different statistics from MLP layers). So, given our limited ability to run operations on the setup, we sampled roughly 10000 dot products from the first ($QK^T$) attention operation and second MLP layer of the first and last encoder layers of the model. The inputs to the whole model were samples from the Wikitext-103 dataset. Our approach captures the range of statistics throughout a model's different components, over its depth, and when processing a real task's data. The second MLP layer has dot product size $4d$, making it the hardest to run experimentally.

In sampling the dot products, we tried to sample from both operands equally. For example, one could sample 1000 dot products by taking a single input vector and 1000 weight matrix vectors, and vice-versa, but choosing random vector pairs captures dot products involving different tokens and weights. This is important because Transformer output sizes, particularly the outlier activation values, are token-dependent \cite{xf_quant}. To maintain this balance, we sample equal rows/columns for both operands. For attention layers we sample 100 from each; For linear layers, we sampled 56 rows from the input data and 200 columns from the weight matrix $W^T$, where the product being computed is $xW^T$.

\section{Simulated Precision Ablation Study}
\label{appendix:ablation}

\begin{table}[]
    \caption{Simulated optical Transformer precision ablation. Input precision is degraded by subsampling from lookup table (LUT), while output is quantized. Input precision is approximate, as LUT has 1000 levels, not 1024. Bold: most compressed model found in our ablation with performance very close to the baseline.}
    \centering
    \begin{tabular}{cc|c}
        \toprule
         Input Precision (LUT) & Output Precision & Val. Loss  \\
         \midrule
         $\sim$ 10 bits & 32 bits & 3.0059 \\
         \midrule
         $\sim$ 9 bits & 32 bits & 3.0057 \\
         $\sim$ 8 bits & 32 bits & 3.0054 \\
         $\sim$ 7 bits & 32 bits & 3.0039 \\
         $\sim$ 6 bits & 32 bits & 3.0034 \\
         $\sim$ 5 bits & 32 bits & 3.0017 \\
         $\sim$ 4 bits & 32 bits & 3.0111 \\
         $\sim$ 3 bits & 32 bits & 3.1223 \\
         $\sim$ 5 bits & 8 bits & 3.0032 \\
         \textbf{$\sim$ 5 bits} & \textbf{7 bits} & \textbf{3.0074} \\
         $\sim$ 5 bits & 6 bits & 3.0335 \\
         $\sim$ 5 bits & 5 bits & 3.3966 \\
         \bottomrule
         
    \end{tabular}
    \label{tab:bit_ablation}
\end{table}

To further study how the optical Transformer can perform inference at lower precisions, we conducted a simple ablation on the input and output precisions used at inference, on the 8-bit-QAT base-sized model with LUT. We opted to leave the weights at 8-bit precision, since in-place weights are not a significant energy cost, and do not take more space/memory in these analog optical systems. In Table \ref{tab:bit_ablation} is the performance of the model at lower precisions. With 5-bit input and 7-bit output precision, the model performs as well as the baseline. The reported precision values for the LUT are approximate, since the LUT has 1000 levels instead of $2^{10} = 1024$ levels.

When using the LUT, it is also not possible to directly change the precision of the input. Instead, we employed a subsampling scheme where the precision is degraded by rounding to every $n$'th integer level before using the LUT, where $n$ is a power of 2 and represents a reduction in the effective bit precision. The LUT of our display has 1000 levels, some levels have the same value, and we simulate the model without added noise. So we say that the original precision is initially \textit{at most} 10 bits ($2^{10} = 1024$).

\section{ONN Energy Calculation}
\label{appendix:energy}

\begin{table*}[h]
    \caption{Designs of models used for energy estimates. Transformers have embedding dimension $d$, process sequence length $n$, use $h$ attention heads, and have $L$ layers. $\mathrm{M}=10^6$ parameters.}
    \centering
    \begin{tabular}{l|llll|l|l}
        \toprule
        Model & n & d & h & L & Parameters & Reference\\
        \midrule
        GPT2 & 1024 & 768 & 12 & 12 & 117M & \cite{gpt2}  \\
        GPT2 & 1024 & 1024 & 16 & 24 & 345M & \\
        GPT2 & 1024 & 1280 & 20 & 36 & 762M &\\
        GPT2 & 1024 & 1600 & 25 & 48 & 1.5B &\\
        Megatron & 2048 & 1536 & 16 & 40 & 1.2B & \cite{megatron-lm}\\
        Megatron & 2048 & 1920 & 20 & 54 & 2.5B &\\
        Megatron & 2048 & 2304 & 24 & 64 & 4.2B &\\
        Megatron & 2048 & 3072 & 32 & 72 & 8.3B &\\
        GPT3 & 2048 & 768 & 12 & 32 & 125M & \cite{gpt3}\\
        GPT3 & 2048 & 1024 & 16 & 24 & 350M &\\
        GPT3 & 2048 & 1536 & 16 & 24 & 760M &\\
        GPT3 & 2048 & 2048 & 24 & 24 & 1.3B &\\
        GPT3 & 2048 & 2560 & 32 & 32 & 2.7B &\\
        GPT3 & 2048 & 4096 & 32 & 32 & 6.7B &\\
        GPT3 & 2048 & 5140 & 40 & 40 & 13B &\\
        GPT3 & 2048 & 12288 & 96 & 96 & 175B &\\
        Turing-NLG & 1024 & 4256 & 28 & 78 & 17B & \cite{t-nlg} \\
        MT-NLG & 2048 & 20480 & 128 & 105 & 530B & \cite{MT-NLG}\\
        Chinchilla & 2048 & 640 & 10 & 10 & 73M & \cite{chinchilla} \\
        Chinchilla & 2048 & 1024 & 16 & 20 & 305M &\\
        Chinchilla & 2048 & 1280 & 10 & 24 & 552M &\\
        Chinchilla & 2048 & 1792 & 14 & 26 & 1.1B &\\
        Chinchilla & 2048 & 2048 & 16 & 28 & 1.6B &\\
        Chinchilla & 2048 & 3584 & 28 & 40 & 6.8B &\\
        Chinchilla & 2048 & 8192 & 64 & 80 & 70B &\\
        PaLM-like & 2048 & 4096 & 16 & 32 & 8B & \cite{palm}\\
        PaLM-like & 2048 & 8192 & 32 & 64 & 62B &\\
        PaLM-like & 2048 & 18432 & 48 & 118 & 540B &\\
        \midrule
        \textbf{FUTURE} & 2048 & 40960 & 80 & 120 & 2.4T & This work\\
        \textbf{FUTURE} & 2048 & 81920 & 128 & 200 & 16T &\\
        \textbf{FUTURE} & 2048 & 163840 & 160 & 400 & 129T &\\
        \textbf{FUTURE} & 2048 & 655360 & 512 & 800 & 4q &\\
        \bottomrule
    \end{tabular}
    \label{tab:energy_models}
\end{table*}

The models we used to estimate the energy use of ONN systems are in Table \ref{tab:energy_models}. We used a variety of real models that have been introduced by other works, and then designed our family of hypothetical future models \textbf{FUTURE-*} in a similar fashion, keeping a reasonable sequence length, increasing the embedding dimension drastically, and following the trend of recent large models like PaLM \cite{palm} and MT-NLG \cite{MT-NLG} of increasing the ratio $d/h$, which results in favorable energy calculations due to the lower fraction of memory operations in attention.

The calculation of energy costs for ONNs requires consideration of the entire system design and the costs of the surrounding electronics---since the optical computation itself is so cheap the electronics account for nearly all of the energy cost. The way the energy is accounted for is as follows: The energy $E_{\mathrm{load}}$ can be broken down into three components, related to the energy of the cost of reading from memory $E_{\mathrm{read}}$, digital-to-analog conversion (DAC) $E_{\mathrm{DAC}}$, and modulation to generate the light $E_{\mathrm{mod}}$: 
\begin{equation}
    E_{\mathrm{load}} = E_{\mathrm{read}} + E_{\mathrm{DAC}} + E_{\mathrm{mod}}.
\end{equation}
Detection energy consumption $E_{\mathrm{det}}$ can broken down in a similar fashion, where 
\begin{equation}
    E_{\mathrm{det}} = E_{\mathrm{amp}} + E_{\mathrm{ADC}} + E_{\mathrm{write}}
\end{equation}
represent the costs of amplifying the detected signal, performing analog-to-digital conversion, and writing to memory respectively. There is also a cost of maintaining the weights in a weights-in-place system, which we call $E_{\mathrm{maintain}}$. Because this cost scales per element, it is a per-MAC cost. But based on values from efficient commercial SLM systems, it is sufficiently small (and amortized by a large clock rate) that even the largest models we do estimations for are not bottlenecked. For optical energy, we take $\SI{1}{\eV}$ (single-photon energy at \SI{1240}{\nano\metre}). We started with using our measured 8-bit-performance photon count of 1500/MAC for the smallest model ($d = 192$) and rescaled the value for larger ones using the constant-per-dot-product trend which we know our simulated models can match or beat.

The assumptions we used were that weights would be loaded from off-chip memory like DRAM (in the case of a chunked-weights strategy; for a full weights-in-place, one-shot approach this cost does not exist), and that the system uses large amounts of SRAM for activations \cite{nvidia_sram}. We assumed that the system only needs 5 bits worth of input precision and 7 bits worth of output precision, per the results of our ablation on the base-sized model. We still assumed 8-bit memory accesses for convenience. The actual costs for the data access and weight maintenance were assumed to be these values: 

\begin{itemize}
    \item $E_{\mathrm{read}}$ = 1 pJ/bit for off-chip memory \cite{sze2017efficient}, and 0.3 pJ/bit for SRAM. The SRAM estimate is based on results for DNN accelerator measurements with 9.55 pJ/32-bit access \cite{acc_sram, acc_sram_2}, and cutting edge/near-future assumptions for data transport from SRAM/cache \cite{nvidia_sram}. \cite{TPUv4i} estimates 14 pJ per 64-bit access, or roughly 0.22 pJ/bit, for a recent TPU architecture.
    
    \item $E_{\mathrm{DAC}}$ = 10 pJ per 5-bit sample @ \SI{10}{\giga\hertz}---this is achievable with ~\SI{100}{\milli\watt} at 30.1dB SFDR \cite{dac_survey}.
    
    \item $E_{\mathrm{mod}}$ = 1 fJ/bit @ \SI{110}{\giga\hertz} with thin-film lithium-niobate modulators \cite{xu2022dual}.
    
    \item $E_{\mathrm{amp}}$ = 2.4 pJ per access. A transimpedance amplifier can run at 24 mW at 70 GHz \cite{tia_100gbps}. We will just assume 10 GHz. 24mW / $10^{10}$ = 2.4 pJ per element.
    
    \item $E_{\mathrm{ADC}} = \SI{3.17}{pJ}$ per 7-bit sample. 10 Ghz, need 7-bits of precision, so 128 conversion steps per sample -- Achievable with 24.8 fJ/c-s \cite{Liu2022} (24.8 fJ $\times$ 128 = 3.17 pJ per 7-bit sample).
    
    \item $E_{\mathrm{write}} = E_{\mathrm{read}}$. Actually, write access was measured to be cheaper than read access in \cite{acc_sram_2}, but we use $E_{\mathrm{write}} = E_{\mathrm{read}}$ as a simple, conservative assumption.
    
    \item $E_{\mathrm{maintain}} = \SI{0.002}{\femto\joule/MAC}$. Assuming 2W for operation of a 10MP SLM, with inputs shone at 10 GHz (each pixel performs one MAC every cycle). There is not much information SLM power consumption for maintenance of a fixed pattern on the LCD panel, though more typical LCD displays which update can operate in the $\sim$1W regime. For example, \cite{sony_lcx} consumes $\SI{30}{\milli\watt}$ with 180000 pixels, which would scale to \SI{1.67}{\watt} with 10MP (at worst, multiple SLMs/LCDs could be used in order to scale up).
\end{itemize}

\section{Breakdown-Of-Costs For Estimated ONN Energy Usage}
\label{appendix:energy_breakdown}
 
\begin{figure}[htb!]
    \centering
    \includegraphics[width=0.4\textwidth]{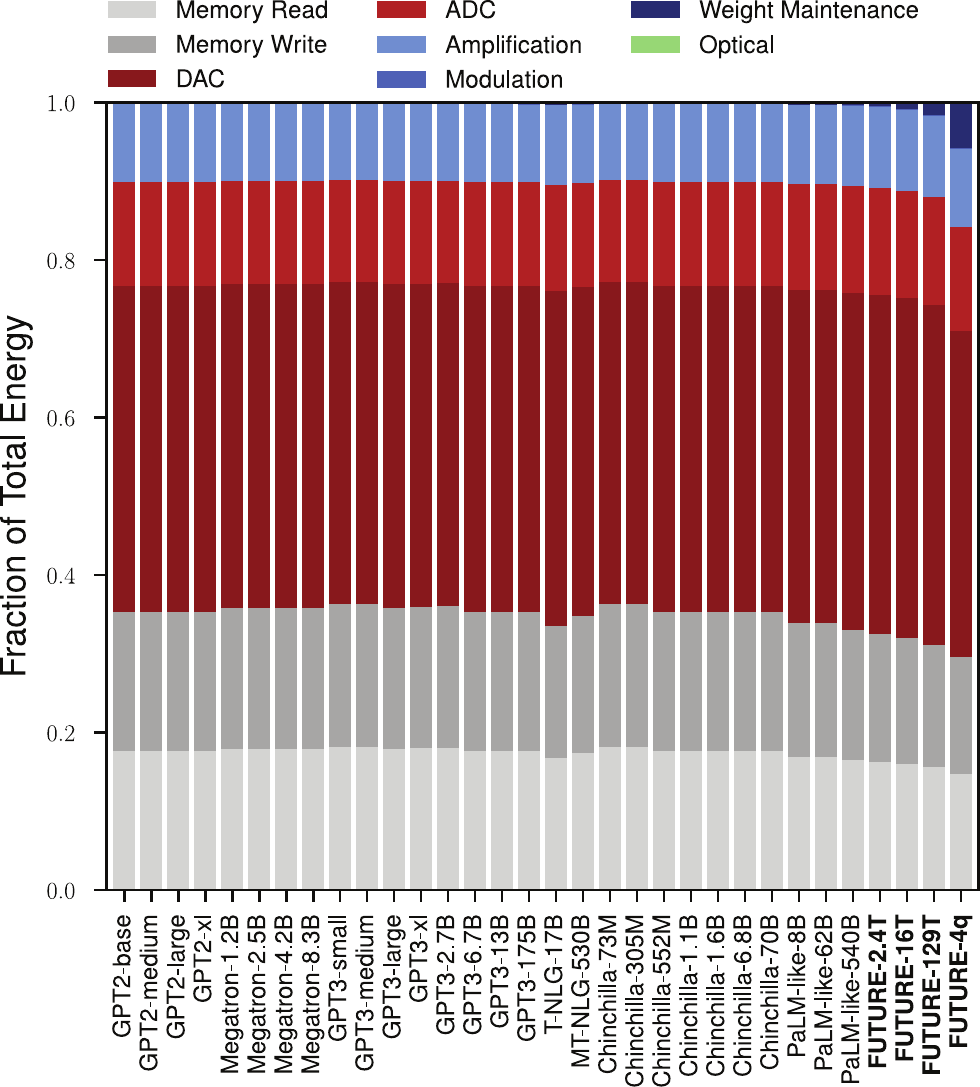}
    \caption{Breakdown of optical Transformer energy costs by energy type at 8-bit operation. Data access costs are dominant due to the high costs of DAC/ADC, but weight maintenance becomes important for large models.}
    \label{fig:energy_type_breakdown}
\end{figure}

 In \cref{fig:energy_type_breakdown} we see that data access costs, that is costs per element loaded/stored in memory, are most expensive. In particular, the cost of ADC and DAC are the leading contributors to the access costs, though since their cost is exponential in the bit precision, one might imagine that a future, optimized Transformer running at lower precision than our 8-bit assumption would have energy costs dominated by the actual SRAM memory costs. Also, for very large models, since the energy from weight maintenance scales with the number of MACs, it eventually will dominate if model sizes scale past that of FUTURE-4q But one might imagine that future hardware would reduce $E_{\mathrm{maintain}}$ through improved electronics or higher clock speeds allowing for lower energy per MAC. Finally, the contribution from optical energy is vanishingly small, a consequence of the efficient photon usage scaling that we found Transformers can leverage. Were it not for this, the cost of actually performing the MACs would be orders of magnitude larger than everything else, resulting in energy usage that scales the same way as digital systems'.

\section{Future ONN Energy Consumption}
\label{appendix:future_energy}

\begin{figure*}[ht!]
    \centering
    \includegraphics[width=0.95\textwidth]{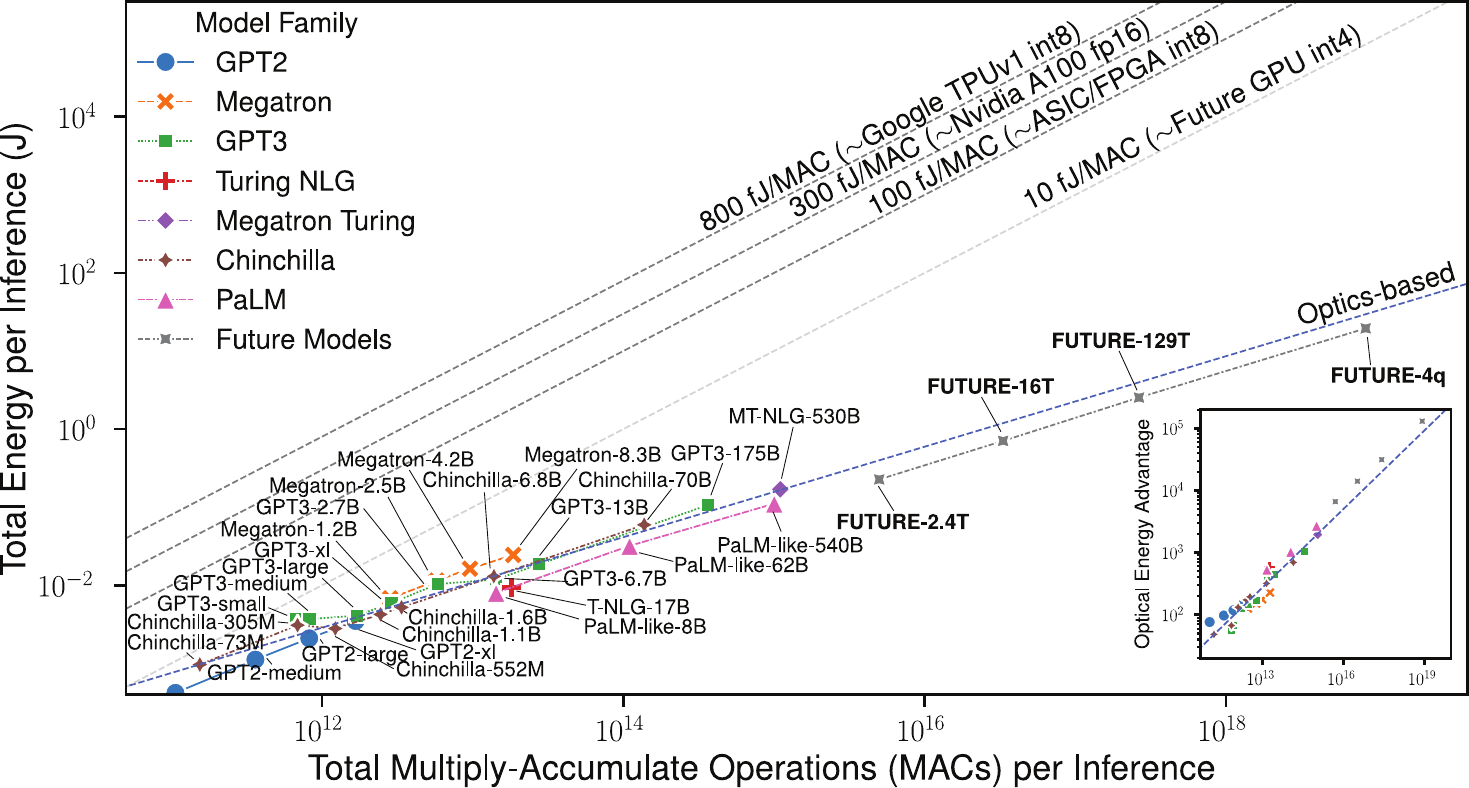}
    \caption{Energy usage estimates of forward pass for Transformers running on optical hardware, under future electronics energy cost assumptions. The energy advantages over our estimate for the current-day NVIDIA A100 GPU are larger than under our original assumptions (\cref{fig:energy}). $\mathrm{M}=10^6$, $\mathrm{G}=10^9$, $\mathrm{T}=10^{12}$, $\mathrm{q}=10^{15}$ parameters.}
    \label{fig:future_energy}
\end{figure*}

As optical accelerators are an emerging technology and as Transformer models continue to scale over time, it is worth considering how ONNs might improve over the next several years. For example, an interesting question to ask is how well future ONNs will do by the time it is possible to run a large model like FUTURE-4q. To investigate this, we estimated the energy costs of various Transformer models running optically again, but with the following changes and assumptions:

\begin{itemize}
    \item $E_{\mathrm{maintain}} = 0$---Future weights-in-place hardware will need effectively no energy to maintain weight information (for example, one might consider the usage of phase change materials \cite{wuttig2017phase}).
    \item $E_{\mathrm{DAC}}$ and $E_{\mathrm{ADC}}$ are 1/32 the size---we assume that electronics could achieve a $2\times$ improvement in fJ/c-s efficiency, while future advancements in model compression allow for 4-bit Transformer models, which are much cheaper since DAC and ADC costs scale exponentially with the number of bits \cite{ADC2020}.
    \item $E_{\mathrm{read}}$ and $E_\mathrm{store}$ are 1/5 the size---there is already a growing recognition of the fact that AI accelerators will need high efficiency and large quantities of SRAM and DRAM in the future \cite{nvidia_sram, cerebras}.
    \item $E_{\mathrm{amp}}$ $10\times$ cheaper (there are already cheaper trans-impedance amplifiers than our conservative estimate here, and receiver-less configuration without any amplifier has also been demonstrated \cite{bandyopadhyay2022single}).
\end{itemize}

Under these assumptions, ONNs become far more efficient, highlighting that improvements to electronics will impact ONNs, and not just competing digital hardware. The energy scaling (\cref{fig:future_energy}) is shifted downward for optics compared to under our previous assumptions, leading to over $1900\times$ and $130,000\times$ advantages over the current A100 GPU for MT-NLG and FUTURE-4q models respectively. Of course, by the time this is possible, GPU efficiency will have improved significantly as well, and we are comparing a 4-bit accelerator to the 16-bit performance of the A100. It is difficult to predict the future efficiency of GPUs at lower precision, but it is clear that ONNs can benefit from improvements to electronics and low-precision inference.

\section{Chunking and Communication Costs in Multi-Processor ONN and GPU Setups}
\label{appendix:chunking}

Implementation of a real ONN for large models might be difficult because the amount of hardware needed to maintain all the weights is exceedingly large. In \cref{table:specs} are the requirements for hardware to run the largest future model. To compute the number of weights/elements, we selected the largest MLP layer in the model, since that requires the most space for weights and activations. While detector and memory requirements are achievable, the number of required cores---each an optical component capable of performing 10M multiplications with weights---is enormous. There are some approaches to remedy this kind of memory issue in both GPUs and ONNs, and we are interested in their hardware-time-energy tradeoffs for ONNs. 

\begin{figure}[h]
    \centering
    \includegraphics[width=0.45\textwidth]{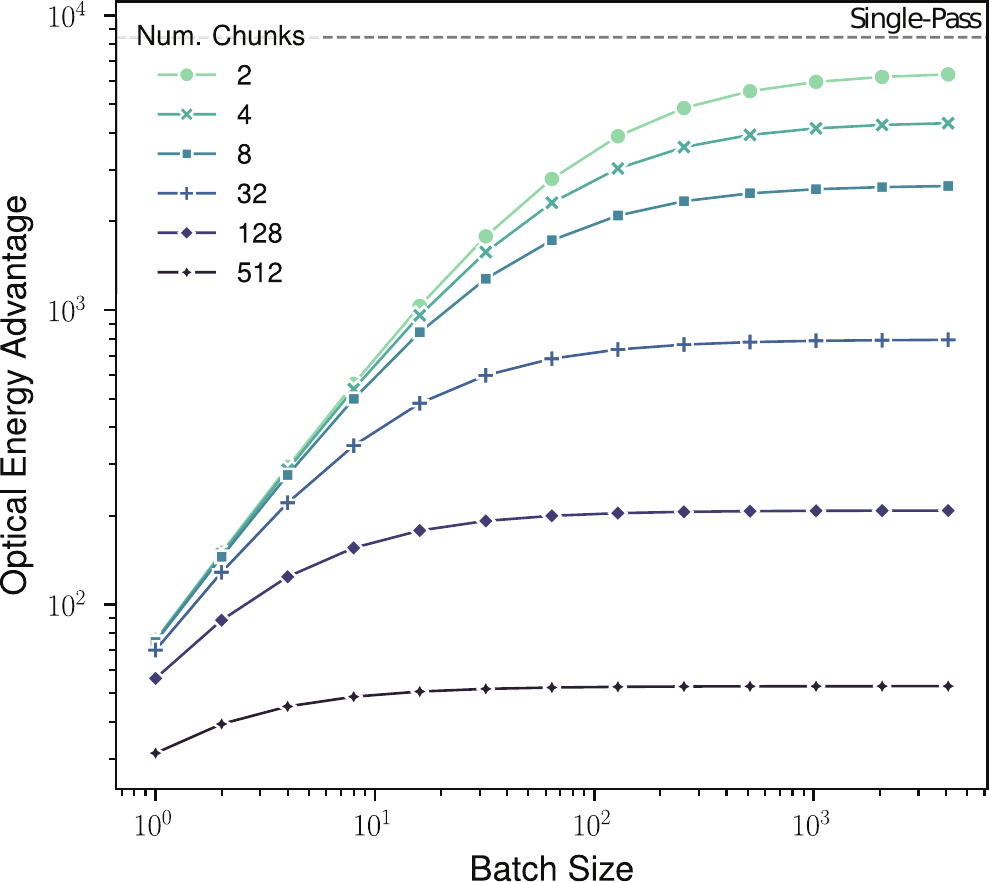}
    \caption{Optical energy advantage vs A100 (FUTURE-4q). When chunking, the cost of loading weights is amortized by increasing batch size, but the overall performance is limited by large numbers of chunks because of input data reloading.}
    \label{fig:chunking}
\end{figure}

One solution is to introduce chunking, where only a portion of the weights are loaded at a time, and the inputs are passed through. Then, the amount of time it takes to run is increased by a factor of the number of chunks. This also impacts the optical system's energy advantage over digital ones in two ways. First, the weights must be loaded, but the cost can be amortized via reuse with batched inference. This comes at the expense of latency. This is a new kind of tradeoff, since digital systems cannot reuse weight data for free. Second, for each weight chunk, all inputs must be reloaded; changing the chunk number trades energy efficiency for lower hardware requirements. These energy tradeoffs are illustrated in \cref{fig:chunking}; other factors dominate energy usage until the chunk number is large and chunking becomes the bottleneck.

Realizing large models with GPUs will likely also  require a multi-GPU strategy, which will incur overhead over the peak performance of a single GPU. We find that with a simple model of communication costs---modelling the activation reloading in both GPU and ONN systems---that ONNs can retain some of their advantages, dependant on how much system memory (or maximum number of weight elements) is available per-processor. We created a simple model to estimate the cost of this approach in GPU systems. In GPU systems, instead of splitting a model over time, the model may instead be split over multiple GPUs. This introduces an analogous tradeoff to the activation reloading in ONNs due to communication costs: if each GPU holds some chunk of weights, then after every layer, the outputs of multiplying the inputs with each chunk must be broadcasted to every GPU in an all-to-all fashion. This is in essence an all-reduce operation---after every layer, the outputs from all GPUs must be copied onto all GPUs. In total, this means the total number of activations is loaded $k$ times, where $k$ is the number of GPUs. As a crude but conservative estimate of these costs, we modeled this by taking the cost of running the entire model on one GPU, and then adding the energy cost of loading the activations from DRAM, multiplied by the number of chunks (GPUs). This is likely an underestimate, as broadcasting data across GPUs in a real setup requires sending data electronically over much longer distances than required for DRAM access, which would be expensive.

To determine the number of chunks, we tested multiple assumptions about device memory. We assumed a value for the amount of memory that can be used to store weights and take the total number of weights for each model divided by this memory capacity to determine the number of chunks to be used.

\begin{figure*}
    \centering
    \includegraphics[width=0.9\textwidth]{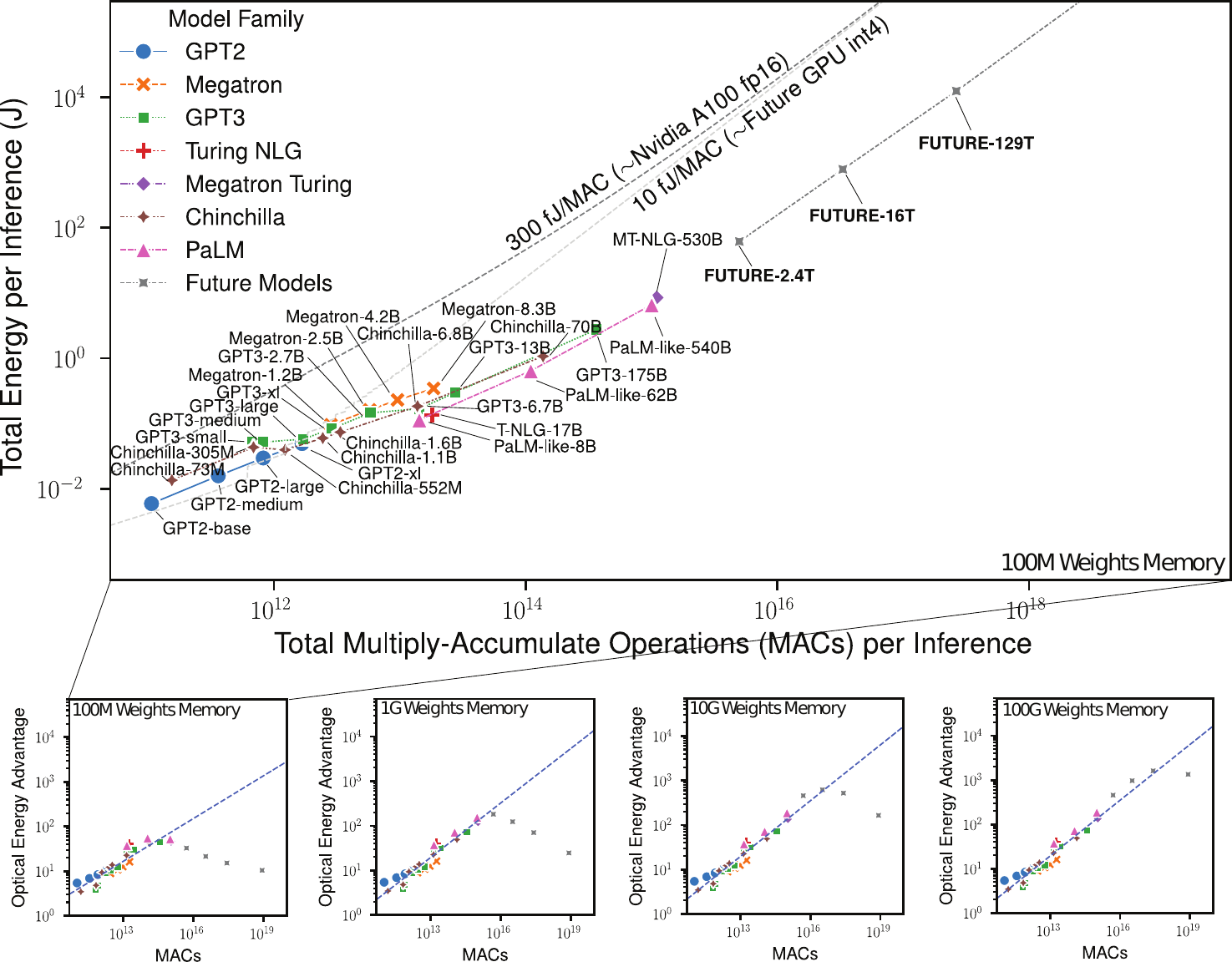}
    \caption{Energy estimates assuming a fixed processor memory size and chunking. Top: estimated energy scaling plot for Transformer models running on optical and digital hardware with 100MB of memory. As models get larger, both optical and digital systems have an upward bend in energy consumption trends, driven by communication/input-reloading-from-chunking costs. Bottom: energy advantage scaling for different memory sizes. As the memory increases, there is a maximum energy advantage for optics over NVIDIA A100 and corresponding model size before chunking costs take over. $\mathrm{M}=10^6$, $\mathrm{G}=10^9$, $\mathrm{T}=10^{12}$, $\mathrm{q}=10^{15}$ parameters.}
    \label{fig:chunking_memory}
\end{figure*}

With these models, we found that too much chunking is detrimental to ONN performance, but that there is still some energy advantage to be had if it is used sparingly (\cref{fig:chunking}). In \cref{fig:chunking_memory} (top) are the energy cost estimates assuming a fixed memory of 100M weights (ie. 100MPixel SLM, or RAM with 100MB capcity if each weight is one byte). We assumed that for GPU, the cost of communication is at least that of DRAM-level communication due to the physical distances between GPUs. The curves for GPUs bend upward as the communication costs begin to take over, as do the largest models running optically. The ONNs still maintain an advantage, but the advantage stops growing with model size. Looking at the energy advantage illustrates this idea more clearly: up to a certain model size the advantage is increasing, then as the model size reaches the memory limit it begins to level off, and then the advantage begins to shrink as the cost of chunking takes over. For a small range of model sizes near this peak, the advantage is maintained, suggesting that a small amount of chunking may be useful before it quickly diminishes the energy advantage. 

The optimal configuration for ONNs, obviously, is to have enough memory (cores which have weights fixed in place) so that chunking is not necessary. When plotting the advantages for larger memories (and therefore fewer chunks), the advantage gets better, and larger models become worthwhile to run. In hindsight this conclusion makes sense: the benefit of ONNs is their ability to copy data (``optical fan-out'') for free for parallel computation, and so reducing this in favor of repeated memory accesses removes exactly the mechanism that gives optics-based systems their advantages. This also suggests that an ``optical memory'' from which fixed data can be accessed for free (or significantly less than re-access through electronics) may solve this problem, allowing for more scalable ONN design without huge amounts of hardware for weights. Currently, optics still has an advantage when using multiple cores because in principle the data could be fanned out across cores, while GPUs must pay communication costs in multi-processor setups. With a fan-out/fan-in design that can collect/spread a vector across cores, the efficiency of an entirely weights-in-place system is fully that of a single, large core.

\begin{table*}[htbp!]
    \caption{Requirements for optical accelerator running feed-forward layer (embedding dimension $d$, sequence length $n$) without chunking at 8-bit precision. The requirement of many cores to maintain weights for matrix-vector products (MVM) is high, and we assume the ONN system requires static RAM (SRAM) for saving and loading activations.}
    \centering
    \begin{tabular}{llllll}
        \toprule
         Model & Input Vector Elements & Detectors & MVM Cores ($10^7$ weights each) & SRAM (activations) \\
         \midrule
         FUTURE-4.1q & $2.6 \times 10^6$ & $2.6 \times 10^6$& 170,000 & 5.37 GB \\
         FUTURE-129T & $6.55 \times 10^5$ & $6.55 \times 10^5$& 11,000 & 1.34 GB \\
         FUTURE-16T & $3.28 \times 10^5$ & $3.28 \times 10^5$& 2,700 & 671 MB \\
         FUTURE-2.4T & $1.64 \times 10^5$ & $1.64 \times 10^5$ & 671 & 336 MB \\
         \midrule
         PaLM-like-540B & $7.37 \times 10^4$ & $7.37 \times 10^4$ & 136 & 151 MB \\
         MT-NLG-530B & $8.19 \times 10^4$ & $8.19 \times 10^4$ & 168 & 168 MB \\
         GPT3-175B & $4.91 \times 10^4$ & $4.91 \times 10^4$ & 61 & 100 MB \\
         \midrule
         \textbf{General} & $4d$ & $4d$ & $4d^2 / 10^{7}$ & $4nd$ \\
         \bottomrule
    \end{tabular}
    \label{table:specs}
\end{table*}

\clearpage
\section{Effects of Training and Quantization Scheme on Optical Scaling}
\label{appendix:clipping}
\begin{figure*}[htbp!]
  \centering
  \includegraphics[width=0.97\textwidth]{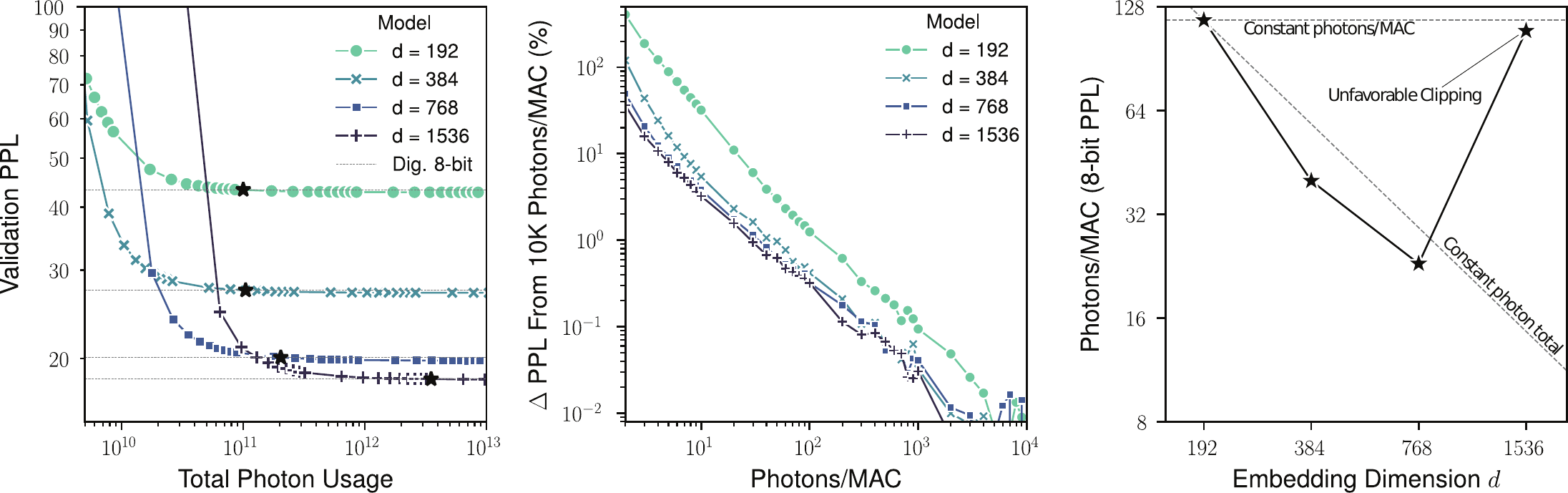}
  \caption{Behavior of optical Transformer models with varying photon usage with percentile clipping scheme. Left: Wikitext-103 validation set perplexity (PPL) versus embedding dimension $d$ and total photons usage. 8-bit quantized digital model performance levels in dashed lines. Middle: Percent change in perplexity from ideal 10000 photon count performance still exhibits truncated power-law scaling with photons per multiply-accumulate (MAC) operation for all models. Right: Scaling of photon usage for maintaining the 8-bit digital performance versus model size. Dashed lines: constant photons per dot product (optical scaling) and constant photons/MAC analogous to digital scaling. Note that unlike for our results in \cref{fig:photonresults}, smaller models beat the constant-dot-product-total scaling, but the largest model exhibits poor efficiency, as the clipping scheme used here was not well suited for it.}
  \label{fig:clippedphotonresults}
\end{figure*}
Our results demonstrating favorable scaling of photon usage in Transformers show that they can be optically efficient, but in general the photon usage is affected by the training scheme and other factors like quantization. This is because approaches for optimization quantization, regularization, etc. affect the statistics of weights and activations in the network, which unlike digital systems, are tied to the resource usage. The main example of this is with weights: they are normalized before being loaded onto an ONN accelerator, and so large outliers may lead to many weights being near 0 after normalization---admitting fewer photons through to the detector. This has a direct impact on the output SNR, and so depending on weight statistics more or fewer photons may be needed in order to run at the same precision.

To discover how a different scheme might affect photon usage, we analyzed the optical scaling of our quantized optical Transformer models with percentile clipping instead of clamping based on EMA statistics. We applied the same clipping to all models (details in Table \ref{tab:quantization}). These clipped models have familiar trends in their language modelling performance versus photon numbers, but we notice key differences in the photons needed to maintain 8-bit digital performance: first, the absolute number of photons needed for the smaller models (120 and 40 versus 340 and 170 of our unclipped scheme for $d=192$, $384$) is much lower---this indicates that clipping of large weight values leads to more transmission after normalization. Second, the scaling is inconsistent, with smaller models needing significantly fewer photons than the expected $1/d$ scaling, but then requiring many photons again for the largest mode. The clipping scheme degraded the performance of the large model. Of course, this could be improved by designing a better scheme for the largest model such that it requires few photons; these results illustrate how differences in the training and quantization recipe could lead to a variety of outcomes, and efficiency is achievable but not an automatic guarantee for any scheme.

\end{appendix}

\end{document}